# A Physics-Driven Framework for Parametric Periodic-Flow Modeling and Finite-Amplitude Aeroelastic Response Analysis

Daiwei Dong[1], Wenbo Cao[2,3], Weiwei Zhang[1,4,5,*]

*1.School of Aeronautic, Northwestern Polytechnical University, Xi'an 710072, China;*

*2. Institute of AI for Industries, Chinese Academy of Sciences, Nanjing 211135, China*

*3. Institute of Computing Technology, Chinese Academy of Sciences, Beijing 100190, China*

*4.International Joint Institute of Artificial Intelligence on Fluid Mechanics, Northwestern Polytechnical University, Xi'an, 710072, China;*

*5.National Key Laboratory of Aircraft Configuration Design, Xi'an 710072, China*

## Abstract

Periodic unsteady flows are common in forced-motion and fluid–structure interaction problems. Their parametric analysis typically requires repeated high-fidelity simulations, whereas existing reduced-order and surrogate models generally rely on pre-generated flow-field or aerodynamic data. This study proposes a purely physics-driven framework for solving parametric periodic flows and finite-amplitude aeroelastic responses. First, a Periodic Physics-Informed Neural Network (P-PINN) is developed to directly solve periodic flows by imposing temporal periodicity over a single motion cycle, thereby avoiding the need to resolve the long transient evolution preceding the establishment of the periodic state. The flow conditions and motion parameters are further incorporated as network inputs to construct continuous parametric representations of the periodic flow field and aerodynamic forces. On this basis, the parametric aerodynamic model is coupled with the structural dynamic equation through first-order harmonic balance to solve the response amplitude and frequency of a single-degree-of-freedom aeroelastic system. This framework is validated using forced-motion cases of a circular cylinder and an airfoil, demonstrating accurate reproduction of periodic aerodynamic forces, surface load distributions, and instantaneous flow fields under different flow and motion parameters compared with time-marching results. Furthermore, aeroelastic analysis is conducted for an elastically mounted circular cylinder at subcritical Reynolds numbers, and the resulting aeroelastic response agrees well with fully coupled CFD/CSD results. Once trained offline, the parametric model can be repeatedly evaluated for different structural parameter states, enabling the complete aeroelastic response curve to be obtained online within seconds, without repeated long-time fluid–structure interaction time marching.

## 1 Introduction

Unsteady flows with moving boundaries are widely encountered in aerospace, marine engineering,

and related fields. When an object undergoes prescribed motions such as plunging and pitching, the resulting motion induces unsteady variations in the surrounding flow field and fluid loads, forming a typical forced-motion problem. When the structural motion is further determined by the interaction between fluid loading and structural dynamics, the problem becomes a fluid–structure interaction problem in aeroelasticity. Such problems are commonly analyzed using high-fidelity CFD or coupled CFD/CSD methods, which typically rely on time marching and must resolve the long transient evolution over multiple cycles before a stable periodic state is established, resulting in considerable computational expense. For forced-motion flows and aeroelastic responses exhibiting stable periodic behavior, periodic steady-state approaches, such as time-spectral [1, 2] and frequency-domain harmonic methods [3-5], exploit temporal periodicity to obtain the periodic state directly over a single cycle, thereby avoiding the preceding transient evolution. However, these methods are primarily formulated for individual parameter states. When the Reynolds number, freestream conditions, motion amplitude and frequency, or structural parameters vary, separate simulations are generally required for different parameter combinations. Consequently, the overall computational cost can still become substantial in large-scale parametric studies as the number of parameter combinations increases.

To reduce the cost associated with repeated high-fidelity simulations at different parameter states, reduced-order modeling and surrogate modeling have become important approaches for unsteady aerodynamic and aeroelastic analysis. Existing methods include reduced-order models based on low-dimensional representations of flow fields, such as proper orthogonal decomposition (POD) [6], as well as input–output models based on Volterra theory [7], autoregressive models with exogenous inputs (ARX) [8, 9], and neural networks[10-13], which establish mappings between flow conditions, structural motions, and unsteady aerodynamic forces or aeroelastic responses. More recently, multi-fidelity learning[14], neural operators[15], and other data-driven models[16] have further enhanced the modeling of complex nonlinear aerodynamic and aeroelastic responses and have been applied to problems such as flutter boundaries, bifurcations, and limit-cycle oscillations. Overall, these reduced-order and surrogate models greatly reduce the online computational cost of aeroelastic analysis. However, their training generally remains dependent on CFD simulations or experimental data, and their accuracy and generalization capability are strongly influenced by the amount of training data, the coverage of the parameter space, and the quality of the available data.

In contrast to these data-driven approaches, physics-informed neural networks (PINNs) provide a physics-driven approach to flow-field solution. By directly incorporating the governing partial differential equations and boundary conditions into the optimization process, PINNs can solve the full flow field without pre-generated flow-field labels. In recent years, PINNs have been widely applied to a range of canonical problems in fluid mechanics. Raissi et al. [17] first demonstrated their potential for inverse problems in fluid flows, while Jin et al. [18] and Rao et al. [19] further established their capability for solving the incompressible Navier–Stokes equations. Subsequent studies extended PINNs to compressible flows [20-24], multiphase flows [25], high-Reynolds-number flows [26, 27], and turbulent flows [28, 29]. To alleviate the training difficulties encountered in more complex flow problems, various improvements have also been developed, including residual weighting [30], variable normalization [31], coordinate transformations [26, 32, 33], and solution strategies [34, 35], further enhancing the applicability of PINNs to forward simulations of complex flows. In addition, by introducing quantities such as Reynolds number, Mach number, angle of attack, and geometric parameters as network inputs, PINNs can establish continuous mappings between physical parameters and flow solutions within a unified model, thereby enabling parametric PDE solutions. This idea has been applied to parametric Navier–Stokes problems and airfoil flows involving variations in freestream conditions and geometry [21, 32, 36, 37], while recent advances in operator learning have further broadened the use of neural networks for parametric PDEs [38].

Compared with steady and parametric steady-flow problems, PINNs remain substantially more challenging for unsteady flows, particularly when long temporal domains, moving boundaries, and continuous parameter variations are involved. To address these difficulties, various strategies has been developed to improve the representation of long-time dynamics, including time-parallel formulations [39], space–time domain decomposition [40], causal training [41-43], and sequential learning [44]. For moving-boundary problems, Farea et al. [45] combined PINNs with an immersed-boundary formulation to model unsteady flows involving moving solid boundaries, although numerical data were still required during training. Zhu et al. [46] directly represented time-dependent moving interfaces and demonstrated physics-driven solutions for several moving-boundary flows without labeled data. More recently, Zou et al. [47] developed a physics-informed graph-based framework for label-free parametric moving-boundary flows, although the investigated parameter ranges were relatively limited. PINNs have also been used for data-assisted reconstruction of moving-boundary flows, where measurements such as PIV velocity data or flow-field data were incorporated to reconstruct pressure and velocity fields [48-50]. These studies demonstrate the potential of PINNs for unsteady and moving-boundary flows. However, constructing label-free parametric models of periodic moving-boundary flows over continuous flow and motion parameter spaces remains insufficiently explored. Furthermore, PINNs have also been introduced into fluid–structure interaction and aeroelasticity[51-54], but existing studies have focused predominantly on data-assisted flow reconstruction, structural-parameter identification, and inverse problems, while fully physics-driven forward solution of aeroelastic responses without labeled data remains relatively underexplored.

Based on the above considerations, this study proposes a physics-driven framework for solving parametric periodic flows and aeroelastic responses with stable periodic behavior. First, a Periodic Physics-Informed Neural Network (P-PINN) is developed by incorporating temporal periodicity directly into the PINNs framework, enabling the periodic unsteady flow to be solved over a single motion cycle. By further introducing flow and motion parameters as network inputs, the P-PINN establishes a parametric model of the periodic flow field and aerodynamic forces over a continuous parameter space. On this basis, the parametric aerodynamic model is coupled with the structural dynamic equation through first-order harmonic balance, transforming the single-degree-of-freedom fluid–structure interaction problem into a low-dimensional nonlinear system with the response amplitude and response frequency as the unknowns, thereby enabling direct solution of finite-amplitude periodic aeroelastic responses. The proposed approach is validated using forced-motion cases of a circular cylinder and an airfoil, and its aeroelastic solution capability is further demonstrated using an elastically supported cylinder. The framework does not rely on pre-generated flow-field or aerodynamic labels, and the trained parametric flow model can be reused for aeroelastic response analysis under different structural parameter states.

The remainder of this paper is organized as follows. Section 2 reviews the governing equations and introduces the basic P-PINN framework, including the temporal periodic boundary condition and the parametric solution strategy. Section 3 validates the capability of P-PINN to solve parametric periodic unsteady flows and the corresponding aerodynamic forces through forced-motion cases of a circular cylinder and an airfoil. Section 4 further combines the parametric aerodynamic model provided by P-PINN with the structural dynamic equation to solve and validate finite-amplitude aeroelastic responses. Finally, Section 5 summarizes the main findings and conclusions of this work.

# 2 Method

## 2.1 Problem setting

This study considers two-dimensional incompressible viscous and inviscid flows. The corresponding incompressible governing equations can be written in a unified form as:

$$
\begin{aligned}
&\nabla \cdot \boldsymbol{u} = 0 \\
&\frac{\partial \boldsymbol{u}}{\partial t} + (\boldsymbol{u} \cdot \nabla)\boldsymbol{u} = -\nabla p + \kappa \nabla^2 \boldsymbol{u},
\end{aligned}
\tag{2.1}
$$

where, $\boldsymbol{u} = (u, v)^T$ is the velocity vector, $p$ is the pressure, and $t$ is the time. The parameter $\kappa$ is introduced to represent the two incompressible flow models considered in this study. When $\kappa = 1/Re$, the governing equations correspond to the incompressible Navier–Stokes equations, where $Re = \rho_\infty U_\infty L_{ref} / \mu$, and $\mu$ is the dynamic viscosity. When $\kappa = 0$, the viscous term vanishes and the governing equations reduce to the incompressible Euler equations. The computational domains, meshes, and boundary conditions for the flows around the circular cylinder and airfoil are illustrated in Figure 1Figure 2, respectively.

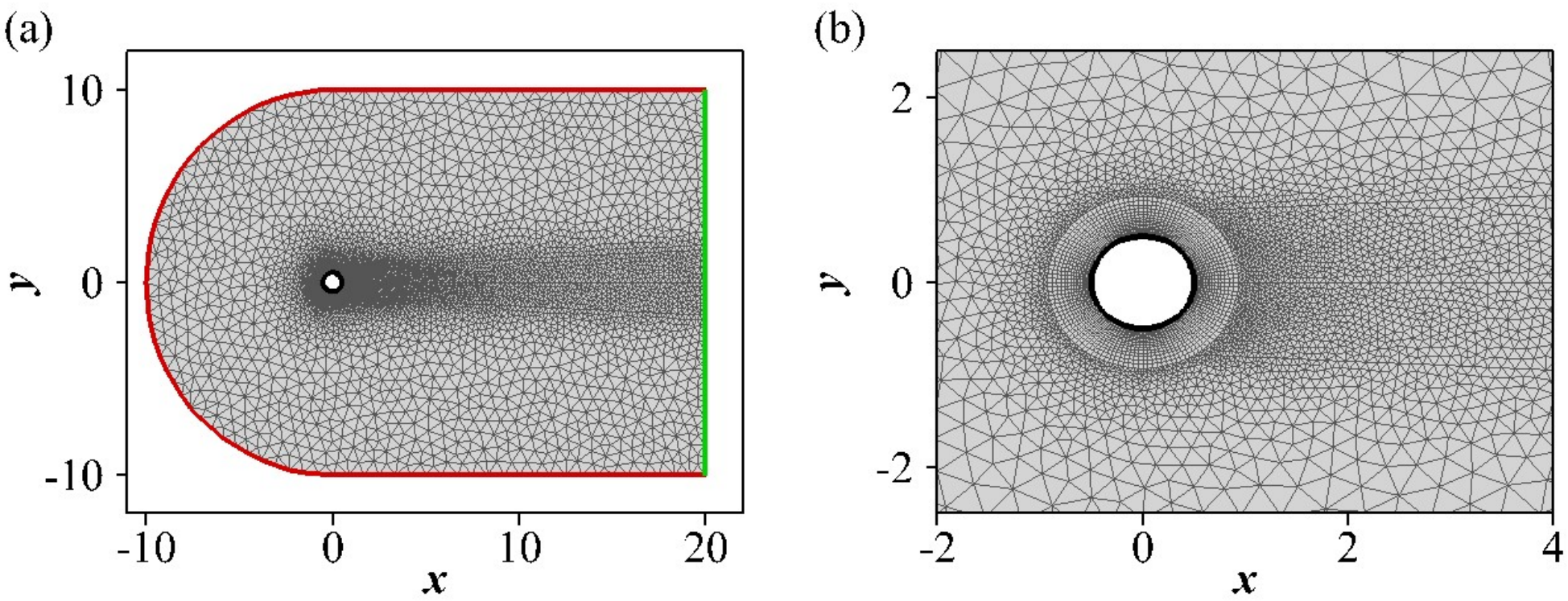

Figure 1. Schematic of the computational domain, mesh, and boundary conditions for the circular cylinder. The red line indicates the velocity inlet boundary, the green line indicates the pressure outlet boundary, and the black line indicates the wall boundary.

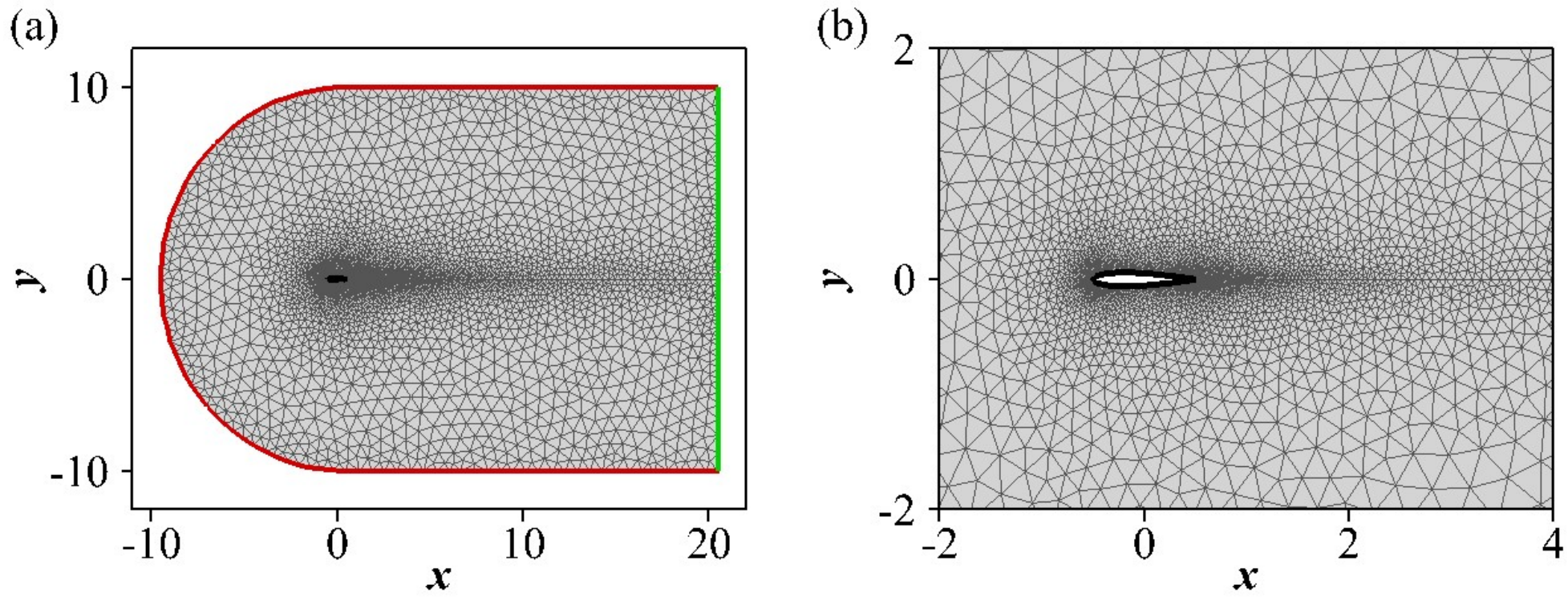

Figure 2. Schematic of the computational domain, mesh, and boundary conditions for the NACA 0012 airfoil. The red line indicates the velocity inlet boundary, the green line indicates the pressure outlet boundary, and the black line indicates the wall boundary.

At the velocity inlet, the uniform freestream condition are $u = cos(\alpha), v = sin(\alpha)$, where $\alpha$ is the angle of attack. At the pressure outlet, the condition is $p = 0$. Different wall boundary conditions are applied to the solid surface according to the governing equations. For the incompressible Navier–Stokes equations, the no-slip condition $\boldsymbol{u} = \boldsymbol{u}_b$ is imposed on the moving wall, where $\boldsymbol{u}_b$ denotes the local velocity of the solid surface. For the incompressible Euler equations, the no-penetration condition $(\boldsymbol{u} - \boldsymbol{u}_b) \cdot \boldsymbol{n} = 0$ is applied, where $\boldsymbol{n}$ is the unit normal vector to the solid wall.

All prescribed motions considered in this study are single-degree-of-freedom sinusoidal motions and can be written in the general form $\eta(t) = A \cdot \sin(2\pi / U^* t)$, where $A$ is the nondimensional motion amplitude, $U^* = U_\infty / fL_{ref}$ is the reduced velocity and $f$ is the motion frequency. For plunging motion, $\eta$ represents the transverse displacement, whereas for pitching motion, $\eta$ denotes the rotation angle about a fixed axis. The pitching axis of the airfoil is located at the quarter-chord position.

## 2.2 Periodic Physics-Informed Neural Network

Physics-informed neural networks (PINNs) employ deep neural networks to approximate the continuous solutions of partial differential equations and train the network parameters by minimizing residual losses associated with the governing equations, boundary conditions, and initial conditions. For the incompressible flows considered in this study, a standard neural network takes the spatial coordinates and time as inputs and predicts the velocity and pressure fields as outputs, which can be written as

$$\hat{\boldsymbol{q}} = \mathcal{N}_\theta(\boldsymbol{x}, t), \qquad \hat{\boldsymbol{q}} = (\hat{u}, \hat{v}, \hat{p})^T, \tag{2.2}$$

where $\theta$ denotes the trainable parameters of the neural network. A typical PINNs loss function is formulated as the mean squared error (MSE) of the residual vector:

$$\mathcal{L} = \frac{1}{N_0} \left\| f(\hat{\boldsymbol{q}}(\cdot;\boldsymbol{\theta})) \right\|_2^2, \tag{2.3}$$

where $N_0$ denotes the dimension of the residual vector $f(\hat{\boldsymbol{q}})$. The residual vector $f(\hat{\boldsymbol{q}})$ consists of the PDE residual $g(\hat{\boldsymbol{q}})$, boundary-condition residual $h(\hat{\boldsymbol{q}})$, and initial-condition residual $i(\hat{\boldsymbol{q}})$, with their relative contributions balanced by the weighting coefficients $\lambda_{PDE}$, $\lambda_{BC}$ and $\lambda_{IC}$ respectively:

$$f(\hat{\boldsymbol{q}}) = \begin{bmatrix} \lambda_{PDE}\, g(\hat{\boldsymbol{q}}) \\ \lambda_{BC}\sqrt{N_g / N_h}\, h(\hat{\boldsymbol{q}}) \\ \lambda_{IC}\sqrt{N_g / N_i}\, i(\hat{\boldsymbol{q}}) \end{bmatrix} = 0, \tag{2.4}$$

where, $N_g$, $N_h$ and $N_i$ denote the dimensions of the PDE residual, boundary-condition residual, and initial-condition residual, respectively. The neural networks predict the solution vector at a set of $m$ collocation points $D = \{\boldsymbol{x}_i, t_i\}_{i=1}^m$, and use automatic differentiation to evaluate the residual vector $f(\hat{\boldsymbol{q}})$. The network parameters are then optimized by minimizing the loss function (2.3) until convergence.

For unsteady flows with stable periodic behavior, conventional PINNs usually need to represent the transient evolution from the initial state to the established periodic state over a long time interval. To avoid this process, we propose a Periodic Physics-Informed Neural Network (P-PINN), which exploits the temporal periodicity of the target flow and transforms the long-time initial-value problem into a temporal boundary-value problem over a single period. In this way, P-PINN directly solves the periodic state without resolving the preceding transient evolution. It should be noted that our previous study [55] showed that, for flows outside the lock-in regime with periodic boundary motion, periodic solutions obtained by optimization-based methods do not necessarily correspond to stable attractors of the original dynamical system. This behavior was demonstrated for a forced oscillating-cylinder flow, where the optimization-based method obtained a non-attracting periodic solution that was difficult to reach through conventional time marching. Therefore, the present study focuses on flow states within the lock-in regime that naturally evolve toward stable periodic responses under time marching.

We define the normalized time as $\tau = t/T$, where $T$ denotes the motion period, so that one fundamental period is mapped to $\tau \in [0,1]$, and the time derivative satisfies $\partial/\partial t = (1/U^*)\partial/\partial\tau$. Unlike conventional PINNs, which impose prescribed initial conditions at the initial time, P-PINN introduces an overlap constraint between the beginning and the end of a single period. Specifically, the corresponding flow states within the two intervals $\tau \in [-\delta_\tau, \delta_\tau]$ and $\tau \in [1-\delta_\tau, 1+\delta_\tau]$ are constrained to be identical, where $\delta_\tau$ denotes the half-width of the temporal overlap region. The corresponding overlap residual vector is defined as:

$$o(\hat{q}) = \hat{\boldsymbol{q}}(x_i, y_i, \tau_i) - \hat{\boldsymbol{q}}(x_i, y_i, \tau_i + 1), \qquad \tau_i \in [-\delta_\tau, \delta_\tau], \tag{2.5}$$

where $N_o$ denotes the number of collocation points in the overlap region. In this study, $\delta_\tau = 0.1$ is selected empirically and is used throughout the remainder of the paper. With this overlap constraint, the flow states at the two ends of the period are matched not only at a single time instant but over finite time intervals, so that temporal periodicity is directly incorporated into the network training. The residual vector of P-PINN is therefore written as:

$$f(\hat{\boldsymbol{q}}) = \begin{bmatrix} \lambda_{PDE} g(\hat{\boldsymbol{q}}) \\ \lambda_{BC}\sqrt{N_g / N_h}\, h(\hat{\boldsymbol{q}}) \\ \lambda_{OC}\sqrt{N_g / N_o}\, o(\hat{\boldsymbol{q}}) \end{bmatrix} = 0. \tag{2.6}$$

To handle the moving boundary, the position of each collocation point $(x_i, y_i, \tau_i)$ in the fluid domain is updated according to the corresponding motion parameters and time phase $\tau_i$. We first calculate the displacement of the solid surface $\Delta\boldsymbol{x}_b$, while the displacement at the far-field boundary is set to zero. We then use a weighted linear interpolation based on the distance from the collocation point to the solid surface $d_b$, and that to the far-field boundary $d_f$, to determine the displacement $\Delta\boldsymbol{x}_i = d_f / (d_b + d_f) \cdot \Delta\boldsymbol{x}_b$ of the collocation point, and finally construct the updated network input $(\tilde{x}_i, \tilde{y}_i, \tau_i)$. In this way, the collocation points on the solid surface move together with the boundary, the far-field boundary remains fixed, and the interior points deform smoothly between them, which preserves smooth and monotonic domain deformation and maintains a regular point distribution.

### 2.3 Time-stepping-oriented neural network and volume-weighted technique

To further improve the training stability and convergence of P-PINN, we incorporate the time-stepping-oriented neural network (TSONN) strategy [34, 35] into the periodic solution framework described above. TSONN introduces a pseudo-time variable independent of the physical time and transforms the original residual equations into a sequence of implicit pseudo-time subproblems:

$$
\begin{aligned}
&PINNs: f(\boldsymbol{q})=0, \\
&TSONN: f(\boldsymbol{q})-(\boldsymbol{q}-\boldsymbol{q}_n)/\Delta s=0, \\
&\boldsymbol{q}_n=\boldsymbol{q}(\cdot;\theta_n), n=0,1,\cdots,N.
\end{aligned} \tag{2.7}
$$

The minimization of the residual of Eq.(2.7) is referred to as the inner iteration. After $K$ inner iterations, we proceed to the next optimization step and update $\boldsymbol{q}_n$ using the latest network output; this process is referred to as the outer iteration. We use the L-BFGS optimizer and set the maximum number of iterations in each optimization step to $K$, which provides a straightforward implementation of the inner iterations. At each outer iteration, we restart the optimizer to accommodate the updated loss function and randomly resample the collocation points.

In addition, we adopt a volume-weighted PDE residual [28, 30] to scale the contributions of densely clustered grid points near the wall. This technique has been shown to perform better for problems with non-uniform grid distributions. Accordingly, the term $g(\boldsymbol{q})$ in Eq.(2.6) is replaced by $\tilde{g}(\boldsymbol{q})$, such that:

$$
\tilde{g}(\boldsymbol{q})=\frac{g(\boldsymbol{q})\boldsymbol{v}}{\sqrt{\frac{1}{N_g}\sum_{i=1}^{N_g} v_i^2}}=\frac{g(\boldsymbol{q})\boldsymbol{v}}{\|\boldsymbol{v}\|_2/\sqrt{N_g}}, \tag{2.8}
$$

where $v_i$ represents the volume occupied by the collocation point $(\boldsymbol{x}_i,\tau)$ in the computational domain. The denominator in the formula represents the root mean square grid volume, which is used for normalization.

**2.4 Parametric Solution of Periodic Flows**

Based on the P-PINN framework described above for periodic-flow solutions, we further introduce flow and motion parameters into the neural-network inputs to enable parametric solutions of periodic flows under different flow conditions and motion states. As illustrated in Figure 3Figure 1Figure 3, the inputs of the parametric solver include the spatial coordinates, normalized time, and the problem-dependent flow and motion parameters.

For the parametric problem, at each outer iteration we randomly sample 30,000 interior residual points, 2,000 boundary points, and 2,000 periodic-overlap points, together with flow and motion parameters randomly selected from the prescribed parameter space. As a result, each collocation point is associated with a different flow condition and motion state, allowing each batch to cover a wide range of parameter combinations. For each collocation point $(x_i, y_i)$, we update its spatial position according to the corresponding time phase $\tau_i$ and motion parameters $(A_i, U_i^*)$ using the moving-boundary treatment described in Section 2.2, thereby obtaining the deformed physical coordinates $(\tilde{x}_i, \tilde{y}_i)$, which are then used to construct the final neural-network input $(\tilde{x}_i, \tilde{y}_i, \tau_i, Re_i, \alpha_i, A_i, U_i^*)$. By resampling both the space–time collocation points and their associated parameter states at every outer iteration, the training process can cover a large number of different parameter combinations and thereby improve the representation and generalization of P-PINN over the continuous parameter space.

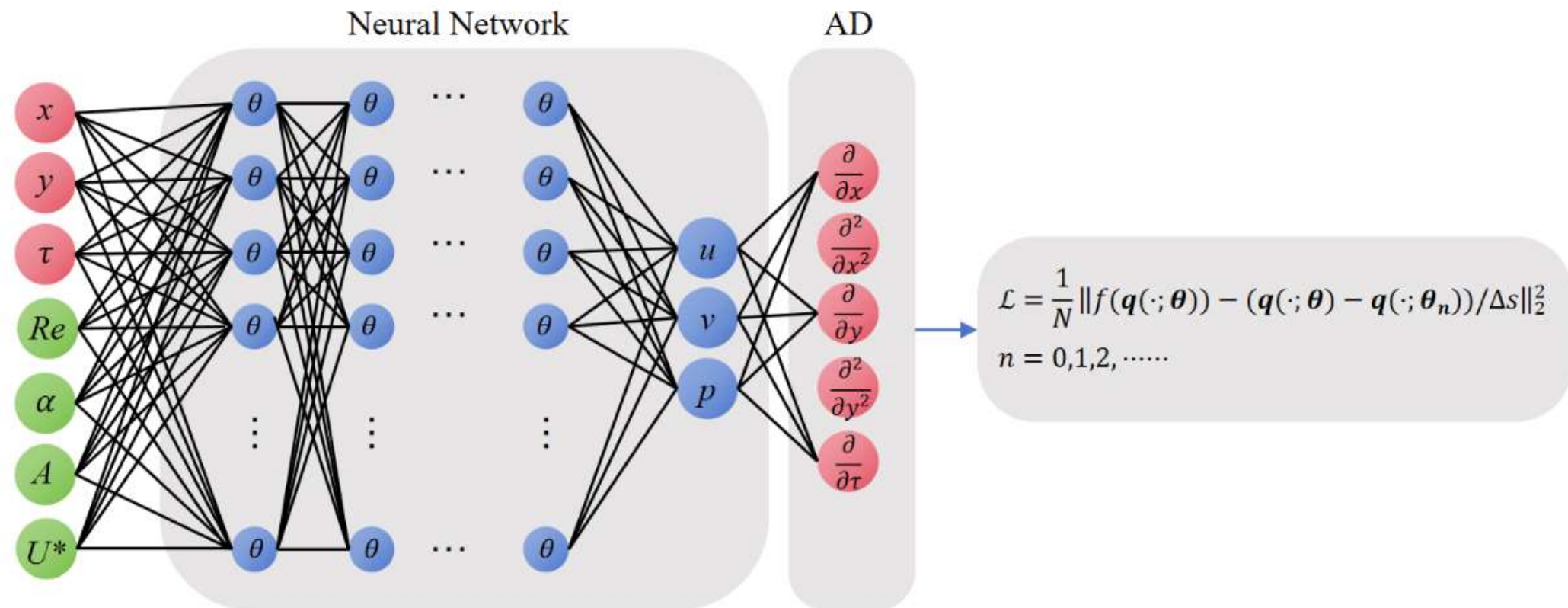

Figure 3. Schematic diagram of the parametric solver based on P-PINN

## 3 Periodic Flow Solutions

This section validates the periodic flow-solving capability of P-PINN through forced-motion cases involving a circular cylinder and an airfoil. The validation is conducted from two perspectives: solving individual flow states and solving over continuous flow and motion parameter spaces. First, the periodic flow field, surface loads, and aerodynamic forces are computed for a forced-plunging cylinder in the lock-in regime at $Re = 100$, and the results are compared with Fluent simulations to assess the accuracy of P-PINN in directly solving periodic unsteady flows. We then consider a forced-plunging cylinder at subcritical and supercritical Reynolds numbers, together with plunging and pitching motions of an airfoil governed by the inviscid equations, to evaluate the parametric solution capability of P-PINN over continuous flow and motion parameter spaces.

To quantitatively assess the differences between P-PINN and the reference results, we use the relative root mean square error (RRMSE) as the primary error metric. For integrated aerodynamic forces, such as the lift coefficient $C_L$ and drag coefficient $C_D$, the error is evaluated at the temporal sampling points over one period:

$$\varepsilon = \sqrt{\frac{\sum_{i=1}^{N_t}(C_i^{\text{P-PINN}} - C_i^{\text{ref}})^2}{\sum_{i=1}^{N_t}(C_i^{\text{ref}})^2}}, \tag{3.1}$$

where, $C^{\text{ref}}$ denotes the reference result, and $N_t$ is the number of temporal sampling points within one period. For surface force distributions such as pressure coefficient $C_p$ and skin-friction coefficient $C_f$, both the temporal sampling points and the discrete points on the body surface are considered, and the error is defined as:

$$\varepsilon = \sqrt{\frac{\sum_{i=1}^{N_t}\sum_{j=1}^{N_s}(C_{i,j}^{\text{P-PINN}} - C_{i,j}^{\text{ref}})^2}{\sum_{i=1}^{N_t}\sum_{j=1}^{N_s}(C_{i,j}^{\text{ref}})^2}}, \tag{3.2}$$

where $N_s$ denotes the number of discrete points on the body surface. Throughout this study, the reference data are obtained from Fluent time-marching simulations, and the periodic results are extracted after the flow has fully developed and reached a stable periodic state.

In all PINN simulations, the L-BFGS optimizer is used with a default learning rate of 1.0. The number of inner iterations, defined as the maximum number of L-BFGS iterations performed during each outer iteration, is set to 250, and the optimizer history size is also set to 250. The relative weights are $\lambda_{PDE} = 100$, $\lambda_{BC} = \lambda_{OC} = 1$, and the pseudo-time step is set to $\Delta s = 0.3$. These parameters are selected based on numerical experience and have also been validated in previous studies [28, 35, 56]. For nonparametric case, the solver uses a fully connected deep neural network with 8 hidden layers, each containing 128 neurons and employing the hyperbolic tangent activation function (tanh). For parametric problems, the number of hidden layers is increased from 8 to 10 to enhance the representational capacity of the model.

### 3.1 Oscillating cylinder

We first validate the capability of the solver to resolve an individual flow state. A single-degree-of-freedom forced-plunging motion of a circular cylinder is considered at $Re = 100$, with the nondimensional motion amplitude and reduced velocity set to $A = 0.25, U^* = 5.5$, respectively. This case lies within a typical lock-in regime, where the cylinder motion and wake vortex shedding develop a stable periodic response. We first compare the overall aerodynamic force response. Figure 4 shows the variations of the lift coefficient $C_L$ and drag coefficient $C_D$ obtained by P-PINN and the reference solution over one complete motion cycle as functions of normalized time.

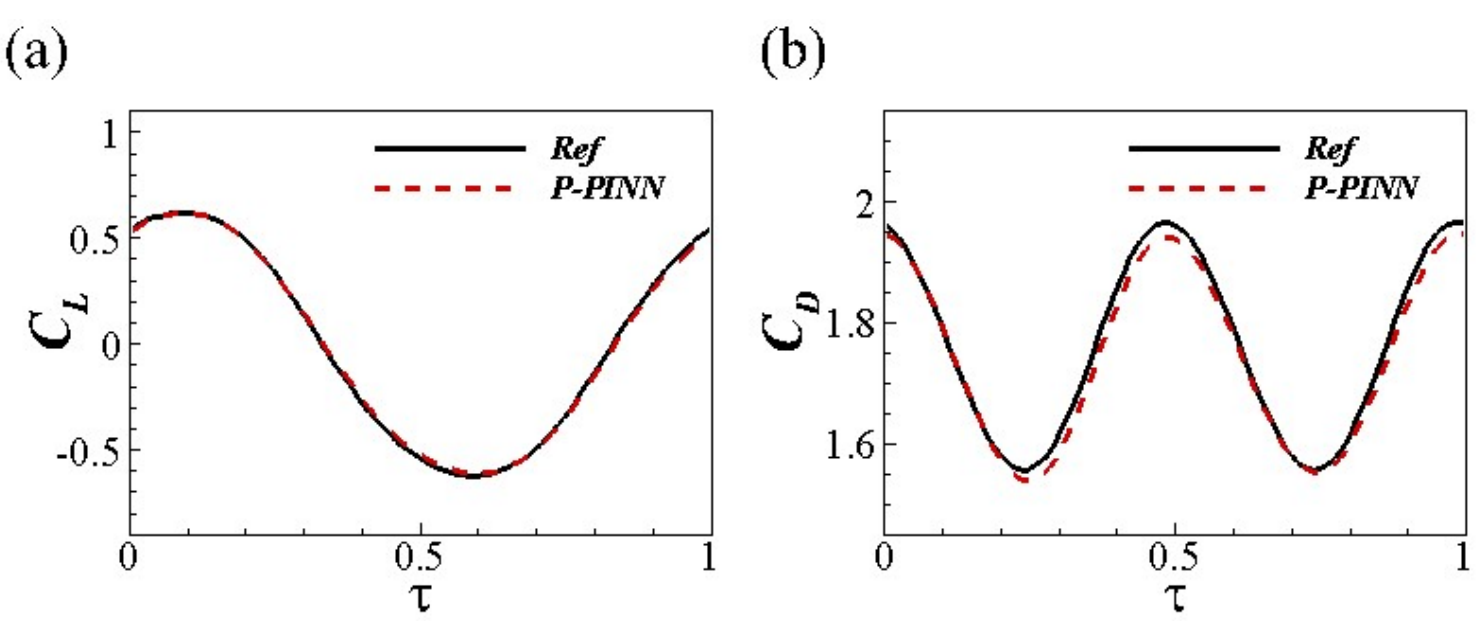

Figure 4. Comparison of aerodynamic coefficients between P-PINN and Fluent at $Re = 100, A = 0.25, U^* = 5.5$: (a) lift coefficient $C_L$; (b) drag coefficient $C_D$.

As shown in Figure 4, the $C_L$ and $C_D$ predicted by P-PINN agree well with the reference results, accurately capturing the periodic variations, amplitudes, and phase characteristics of the aerodynamic forces. Their relative root mean square errors are 0.03076 and 0.01066, respectively, indicating a high level of accuracy. To further examine the local load distributions at different phases within one motion cycle, four representative phases close to $\tau = 0$, 0.25, 0.5 and 0.75 are selected for comparison. Because Fluent advances the solution using physical time steps, the actual output times do not exactly coincide with the phases listed above, and the phases extracted from the P-PINN results are chosen to be consistent with those of the Fluent results. Figure 5 presents the distributions of the pressure coefficient $C_p$ and skin-friction coefficient $C_f$, together with the corresponding errors, at the representative phases. The exact phase values are indicated in the figure.

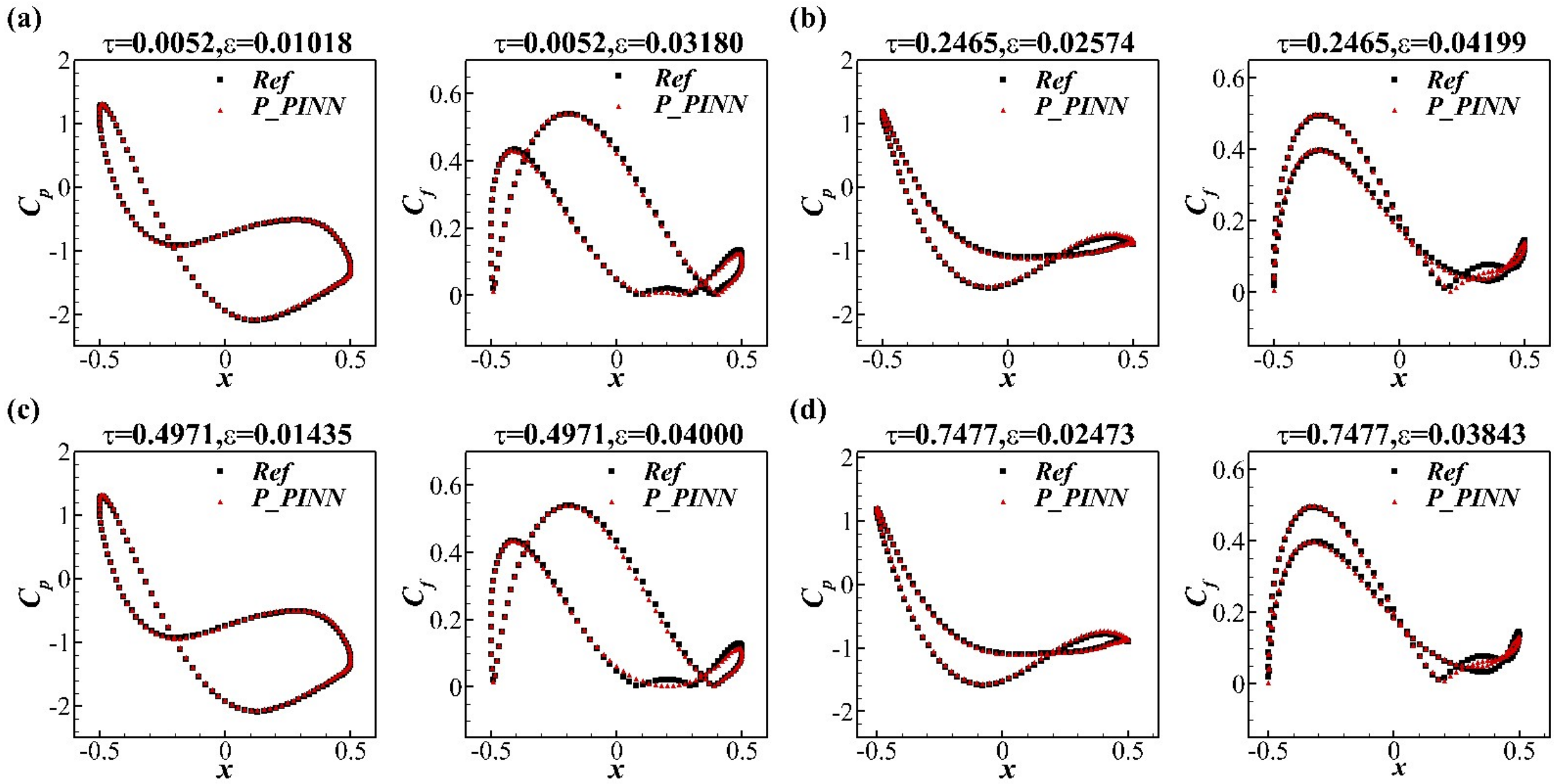

Figure 5. Comparison of the surface pressure coefficient $C_p$ and skin-friction coefficient $C_f$ between P-PINN and Fluent at four representative phases: (a) $\tau \approx 0$; (b) $\tau \approx 0.25$; (c) $\tau \approx 0.5$; (d) $\tau \approx 0.75$.

As shown in Figure 5, the predicted $C_p$ and $C_f$ agree closely with the reference results at all phases, with an RRMSE of 0.01803 and 0.04049 over the entire cycle. P-PINN accurately describes the variations in the cylinder-surface pressure and wall shear stress throughout the motion cycle.

Finally, the instantaneous flow-field solution is further validated. To avoid repeatedly presenting flow-field results at different phases, the instant $\tau = 0.0052$ is selected as a representative time for detailed comparison. Figure 6 shows the velocity components $u$ and $v$, pressure $p$, and vorticity magnitude $|\omega|$ obtained by P-PINN and Fluent, together with the corresponding pointwise error distributions. P-PINN accurately captures the main flow structures near the cylinder surface and in the wake, particularly the formation and shedding of wake vortices. The predicted velocity, pressure, and vorticity fields agree well with the Fluent results, and the pointwise errors of all physical quantities remain relatively small. Combined with the preceding comparisons of the periodic aerodynamic forces and surface loads, these results demonstrate that P-PINN can accurately resolve the periodic aerodynamic forces, surface loads, and instantaneous flow field of the cylinder in the lock-in regime, thereby validating its capability to solve periodic unsteady flows with moving boundaries.

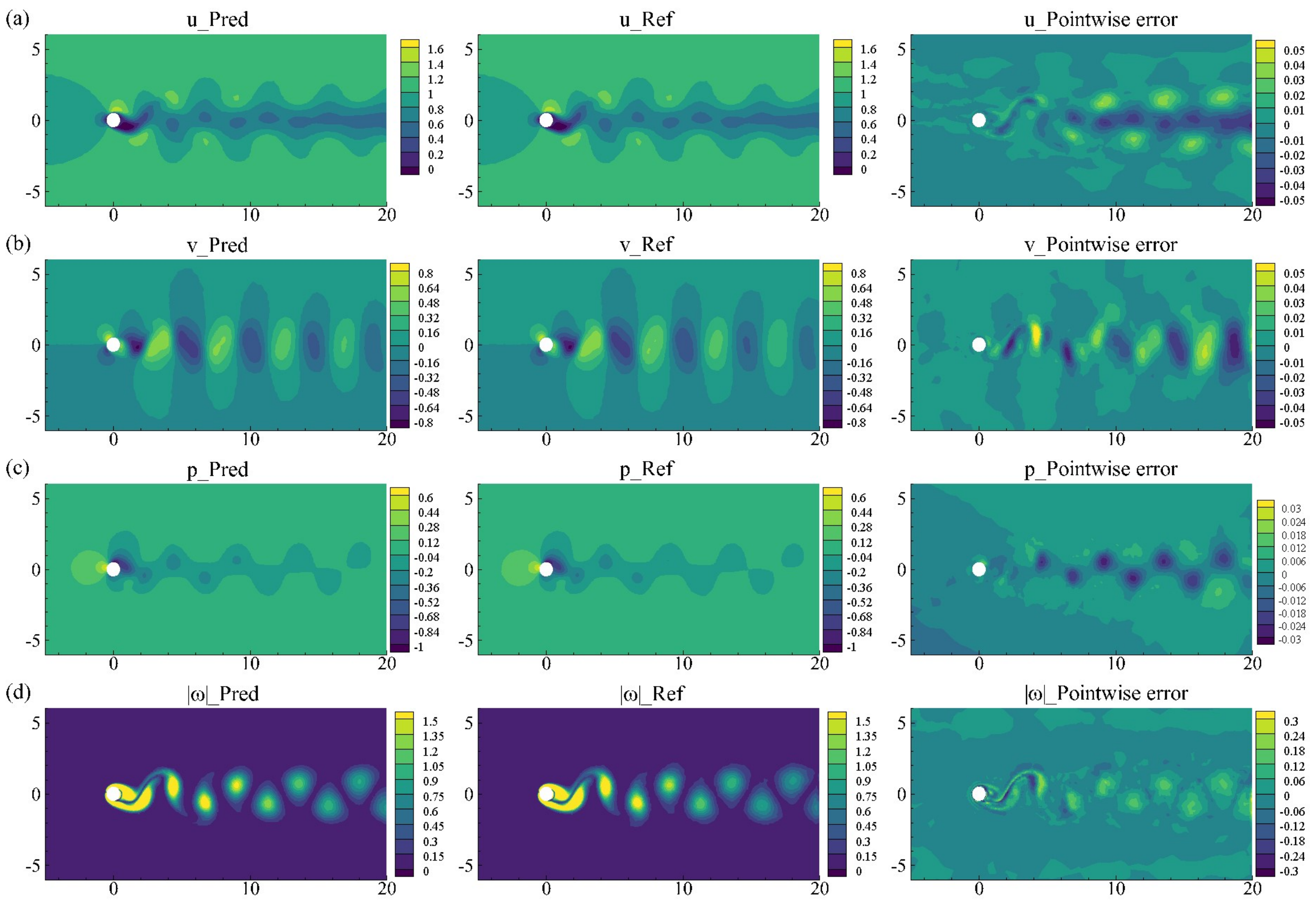

Figure 6. Comparison of the instantaneous flow fields between P-PINN and Fluent at phases $\tau = 0.0052$, together with the corresponding pointwise errors: (a) $u$; (b) $v$; (c) $p$; (d) vorticity magnitude $|\omega|$.

### 3.2 Parametric Oscillating Cylinder

Building on the validation of an individual flow state, we further investigate the capability of P-PINN to solve periodic cylinder flows parametrically under different flow and motion parameters. Here, the terms subcritical and supercritical Reynolds numbers refer to two types of cases relative to the critical state for natural vortex shedding from a stationary cylinder, and all validation cases are selected within the corresponding parameter ranges. Because the approximation accuracy of a parametric neural network may deteriorate near the boundaries of the parameter domain, the actual training ranges are appropriately extended beyond the target solution intervals, and a certain overlap is retained near the critical Reynolds number to reduce the influence of parameter-domain boundaries on the solution accuracy.

#### 3.2.1 Subcritical Reynolds Number Flow

We first validate the lock-in states within the subcritical Reynolds-number range, with the parameter ranges set to $Re \in [5,50], A \in [0.05,0.30], U^* \in [2,7]$. Table 1 lists the parameter combinations of the four validation cases, together with the RRMSEs of the lift coefficient, drag coefficient, and surface-distributed quantities over one motion cycle. Figure 7 shows the periodic variations of $C_L$ and $C_D$ for the four validation cases. Considering the symmetry of the cylinder-plunging motion, two representative phases close to $\tau = 0$ and 0.25 are selected to compare the distributions of the $C_p$ and $C_f$, and the results are shown in Figure 8.

Table 1. Parameter settings of the validation cases and corresponding RRMSE of $C_L$, $C_D$, $C_p$ and $C_f$

| | ***Re*** | ***A*** | ***U**** | $\varepsilon_C_L$ | $\varepsilon_C_D$ | $\varepsilon_C_p$ | $\varepsilon_C_f$ |
|---|---|---|---|---|---|---|---|
| **Case 1** | 44 | 0.15 | 4 | 0.0239 | 0.0041 | 0.0523 | 0.0267 |
| **Case 2** | 33 | 0.25 | 5 | 0.0160 | 0.0114 | 0.0624 | 0.0287 |
| **Case 3** | 26 | 0.10 | 6 | 0.0484 | 0.0034 | 0.0284 | 0.0195 |
| **Case 4** | 15 | 0.15 | 3 | 0.0489 | 0.0066 | 0.1511 | 0.0177 |

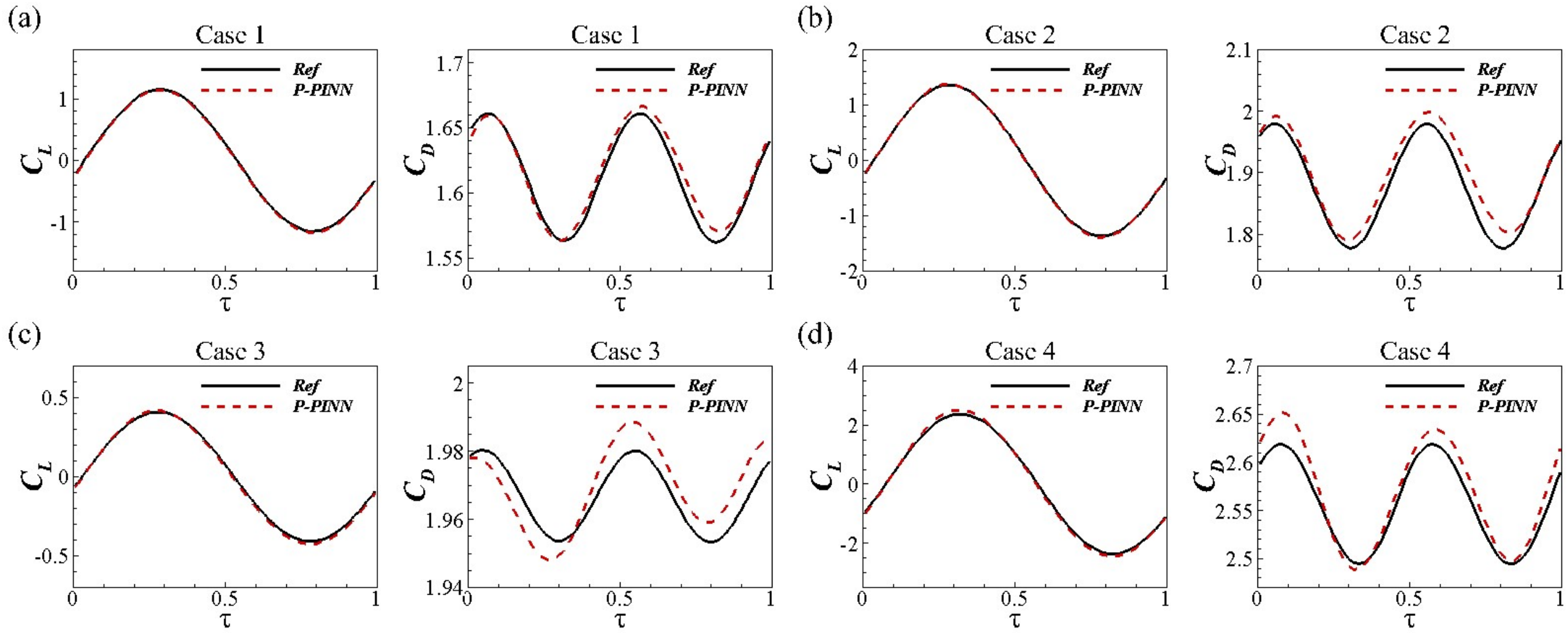

Figure 7. Comparison of the $C_L$ and $C_D$ between P-PINN and Fluent for the four validation cases: (a) Case 1; (b) Case 2; (c) Case 3; (d) Case 4.

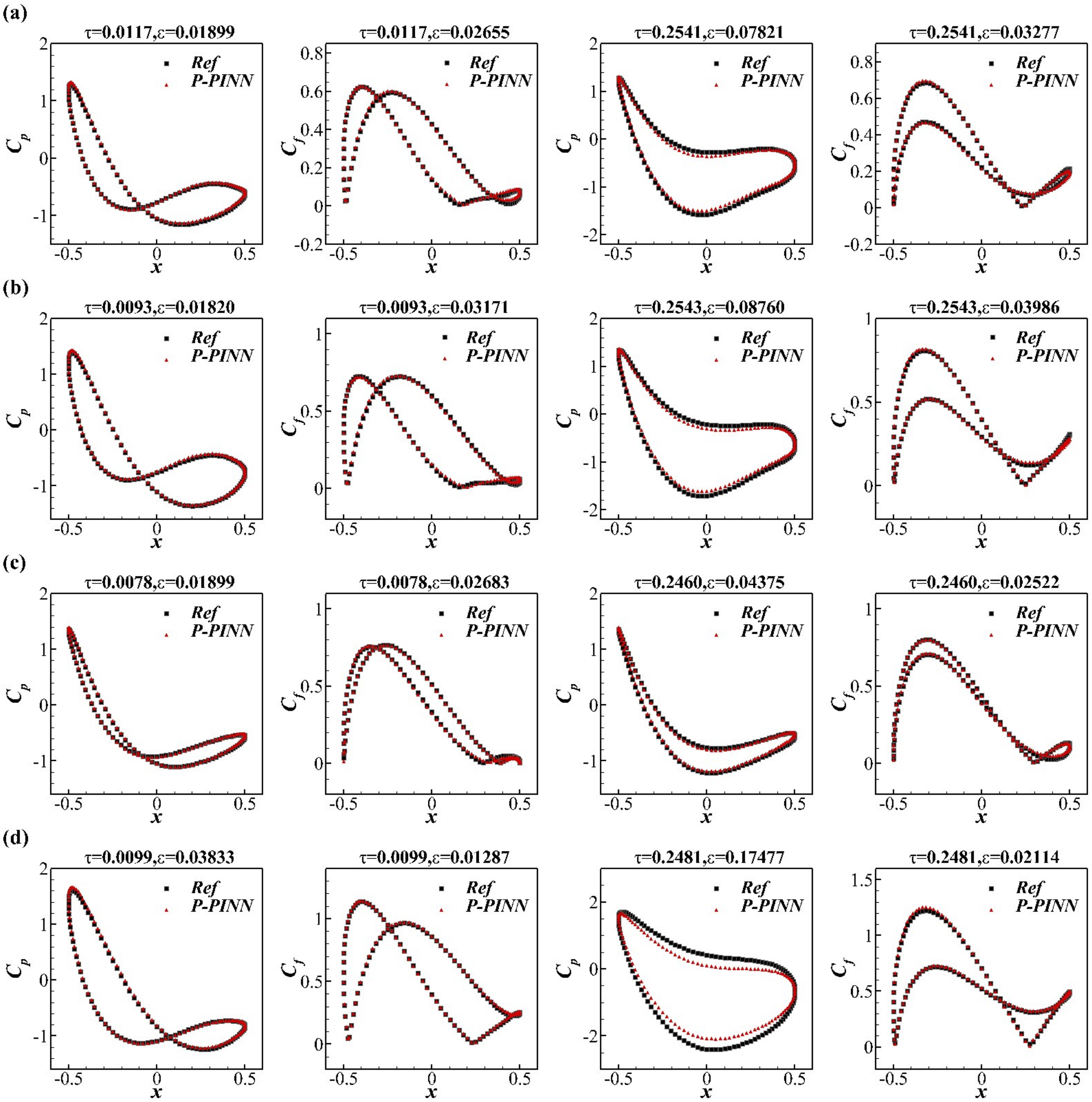

Figure 8. Comparison of $C_p$ and $C_f$ between P-PINN and Fluent for the four validation cases at two representative phases, $\tau \approx 0$ and $\tau \approx 0.25$: (a) Case 1; (b) Case 2; (c) Case 3; (d) Case 4.

The results in Table 1 and Figure 7Figure 8 show that P-PINN maintains high accuracy in predicting the periodic aerodynamic forces for all four validation cases. The predicted $C_L$ and $C_D$ agree well with the Fluent results in terms of amplitude, phase, and overall variation trend. For the cylinder-surface loads, both $C_p$ and $C_f$ maintain high accuracy across the validation cases. In particular, the local distribution of $C_f$ near the rear of the cylinder is also accurately reproduced. It should be noted that the RRMSE of $C_p$ for Case 4 is 0.1511, which is slightly larger than those of the other cases. The main error occurs near the phase at which the cylinder reaches its maximum displacement. At this phase, the $C_p$ distributions near the leading and trailing sides of the cylinder remain in good agreement with the Fluent results, whereas a certain amplitude deviation appears over the middle part of the cylinder surface, increasing the overall RRMSE for this case. Nevertheless, the overall distribution trend of $C_p$ is still accurately captured, and the RRMSE of $C_L$ is only 0.0489, indicating that this local deviation does not significantly affect the accuracy of the integrated aerodynamic force.

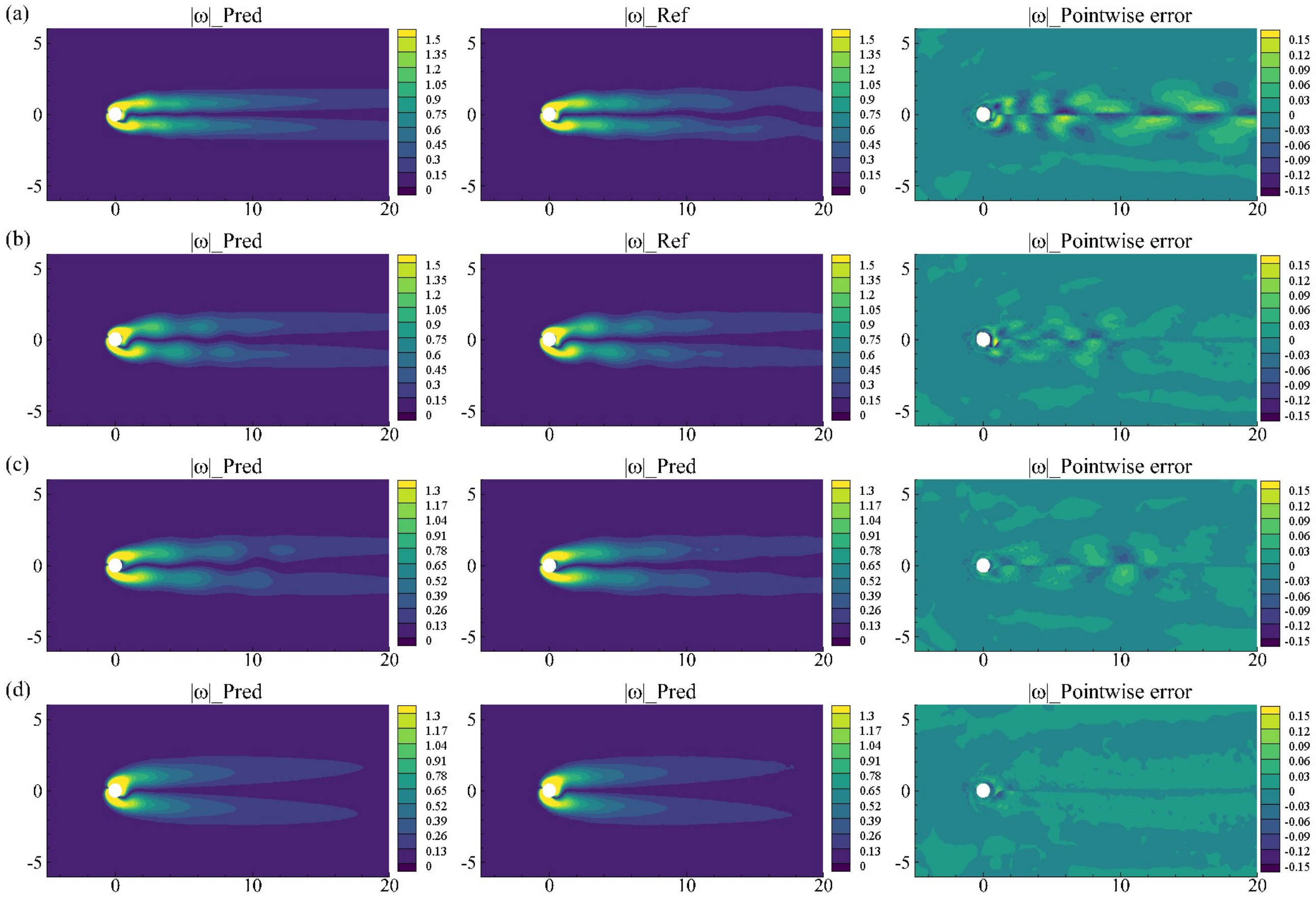

Figure 9. Comparison of the vorticity magnitude $|\omega|$ between P-PINN and Fluent at phases $\tau \approx 0$ for the four validation cases, together with the corresponding pointwise errors: (a) Case 1; (b) Case 2; (c) Case 3; (d) Case 4.

To further examine the capability of the parametric P-PINN to represent wake-flow structures, Figure 9 compares the instantaneous vorticity fields of the four validation cases at a phase close to $\tau = 0$ with the corresponding Fluent results. Overall, P-PINN accurately captures the main vortex structures near the cylinder surface and in the wake under different parameter conditions. The locations, shapes, and intensities of the vorticity distributions are generally consistent with the reference results. It should be noted that the wake structures in the downstream region are described with some deviation for Case 1, whereas high accuracy is maintained near the cylinder surface. To further analyze this difference, Figure 10 compares the velocity components $u$ and $v$, together with the pressure $p$, for Case 1 at the same instant. The results show that the main deviations are concentrated in the far-wake region. In this case, the wake vortices are relatively weak, and the variations in $u$ and $v$ in the wake are small. Although the absolute errors between P-PINN and Fluent remain relatively low, they become more pronounced in the vorticity distribution because of the weak vorticity magnitude in this region.

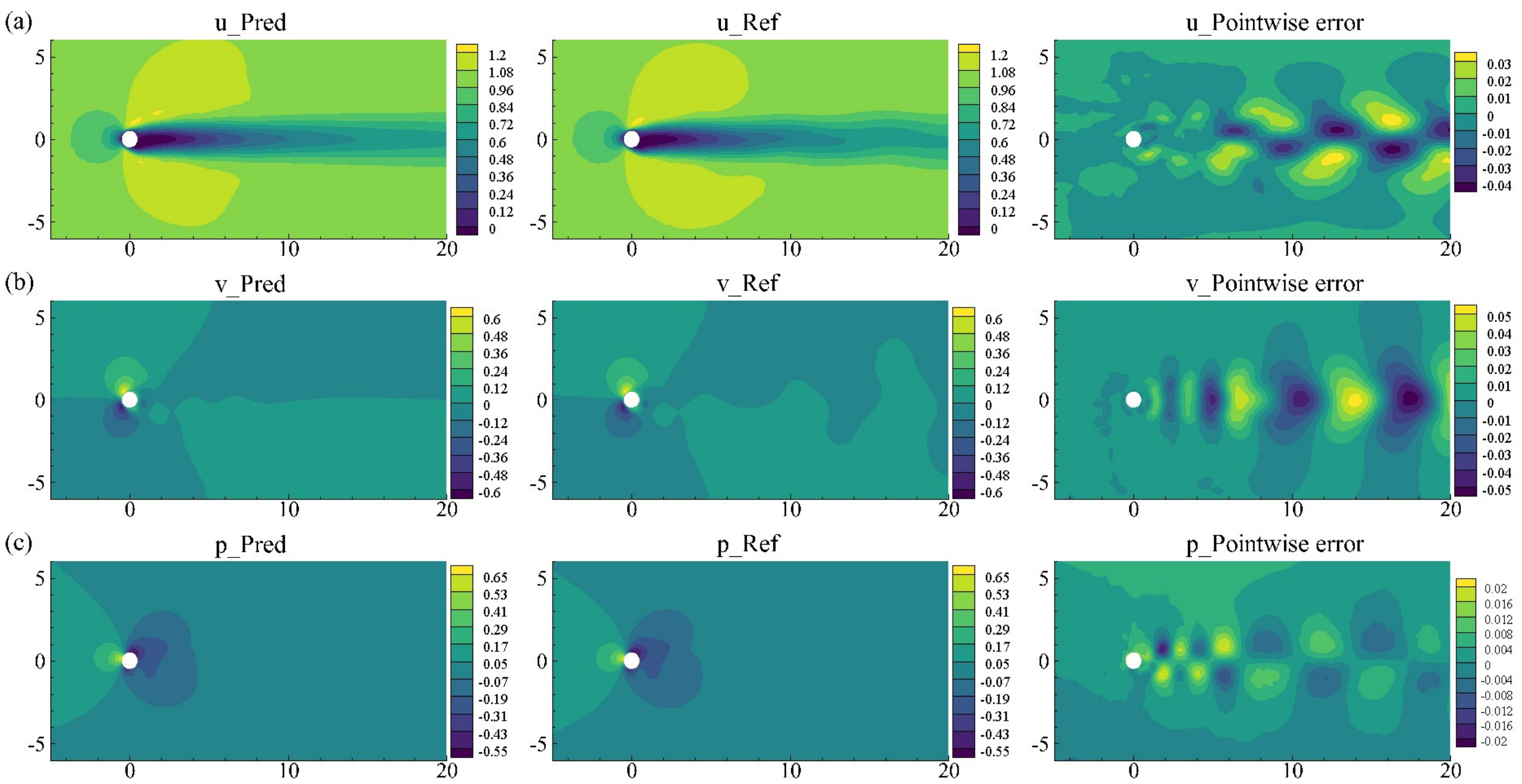

Figure 10. Comparison of the instantaneous velocity and pressure fields between P-PINN and Fluent at phases $\tau \approx 0$ for Case 1, together with the corresponding pointwise errors: (a) $u$ ; (b) $v$ ; (c) $p$ .

Based on this observation, we further select lock-in states at supercritical Reynolds numbers for parametric validation to examine the capability of P-PINN to represent stronger wake-vortex structures. In these cases, vortex shedding in the wake is more pronounced, providing an opportunity to determine whether the wake deviation observed in the subcritical cases originates from limitations in P-PINN's ability to represent the vortex structures themselves.

### 3.2.2 Supercritical Reynolds Number Flow

Building on the preceding parametric solution of subcritical Reynolds-number flows, we further investigate the capability of P-PINN to solve lock-in states at supercritical Reynolds numbers. The parameter ranges of the corresponding parametric model are set to $Re \in [40,120], A \in [0.05, 0.30]$, $U^* \in [5,9]$. Table 2 lists the parameter combinations of the four validation cases, together with the RRMSEs of the lift coefficient, drag coefficient, and surface-distributed quantities over one motion cycle. Figure 11 shows the periodic variations of $C_L$ and $C_D$ for the four validation cases. Similarly, considering the symmetry of the cylinder-plunging motion, two representative phases close to $\tau = 0$ and 0.25 are selected to compare the distributions of $C_p$ and $C_f$ over the cylinder surface, as shown in Figure 12.

Table 2. Parameter settings of the validation cases and corresponding RRMSE of $C_L$, $C_D$, $C_p$ and $C_f$

| | ***Re*** | ***A*** | ***U**** | $\varepsilon_C_L$ | $\varepsilon_C_D$ | $\varepsilon_C_p$ | $\varepsilon_C_f$ |
|---|---|---|---|---|---|---|---|
| **Case 1** | 60 | 0.15 | 7 | 0.1782 | 0.0051 | 0.0129 | 0.0173 |
| **Case 2** | 80 | 0.25 | 7.5 | 0.2627 | 0.0089 | 0.0161 | 0.0190 |
| **Case 3** | 100 | 0.10 | 6.5 | 0.1353 | 0.0033 | 0.0124 | 0.0219 |
| **Case 4** | 100 | 0.25 | 5.5 | 0.0987 | 0.0120 | 0.0328 | 0.0518 |

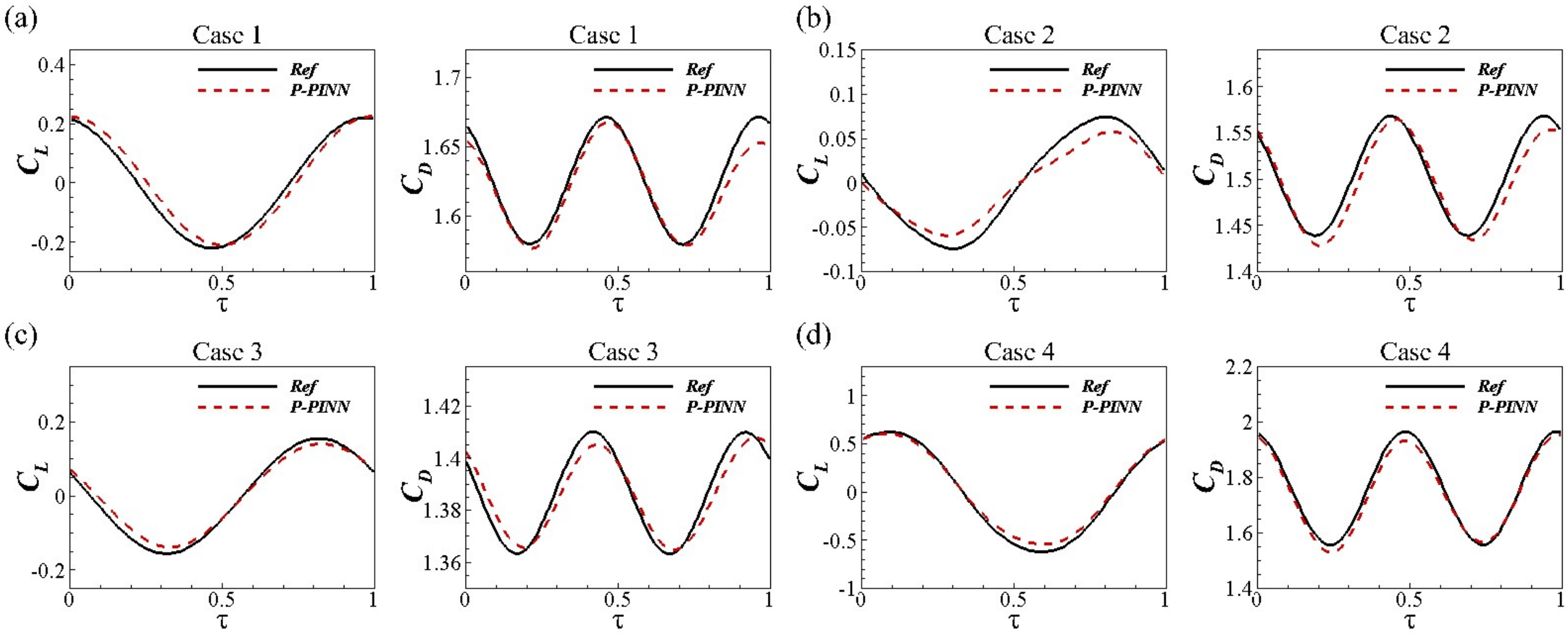

Figure 11. Comparison of the $C_L$ and $C_D$ between P-PINN and Fluent for the four validation cases: (a) Case 1; (b) Case 2; (c) Case 3; (d) Case 4.

The results show that P-PINN achieves high accuracy in predicting the $C_D$ and surface loads at supercritical Reynolds numbers. For all validation cases, the $C_p$ and $C_f$ distributions obtained by P-PINN accurately reproduce the reference results over the entire cylinder surface at different representative phases within one motion cycle, including regions with large pressure gradients and the local surface loads near the rear of the cylinder. This indicates that the model can accurately describe the near-wall flow and its variation with the motion phase under different parameter conditions.

In comparison, the RRMSE of $C_L$ is relatively large. It should be noted that this does not indicate a general reduction in the accuracy of the cylinder-surface loads. In these cases, the pressure and viscous-shear contributions from different surface regions in the lift direction strongly cancel each other, resulting in a relatively small net lift amplitude. Consequently, even when the local errors in $C_p$ and $C_f$ over the entire surface are small, these small differences may produce a relatively large error in the integrated $C_L$ after cancellation and integration. Moreover, because the amplitude of $C_L$ itself is small, the RRMSE normalized by the mean-square reference value further amplifies this difference, leading to a relatively large error value. Therefore, the relatively large relative error in $C_L$ mainly reflects the sensitivity of a small-amplitude integrated quantity to local errors, whereas P-PINN still maintains high accuracy in the complete surface-pressure and shear-stress distributions contributing to this quantity.

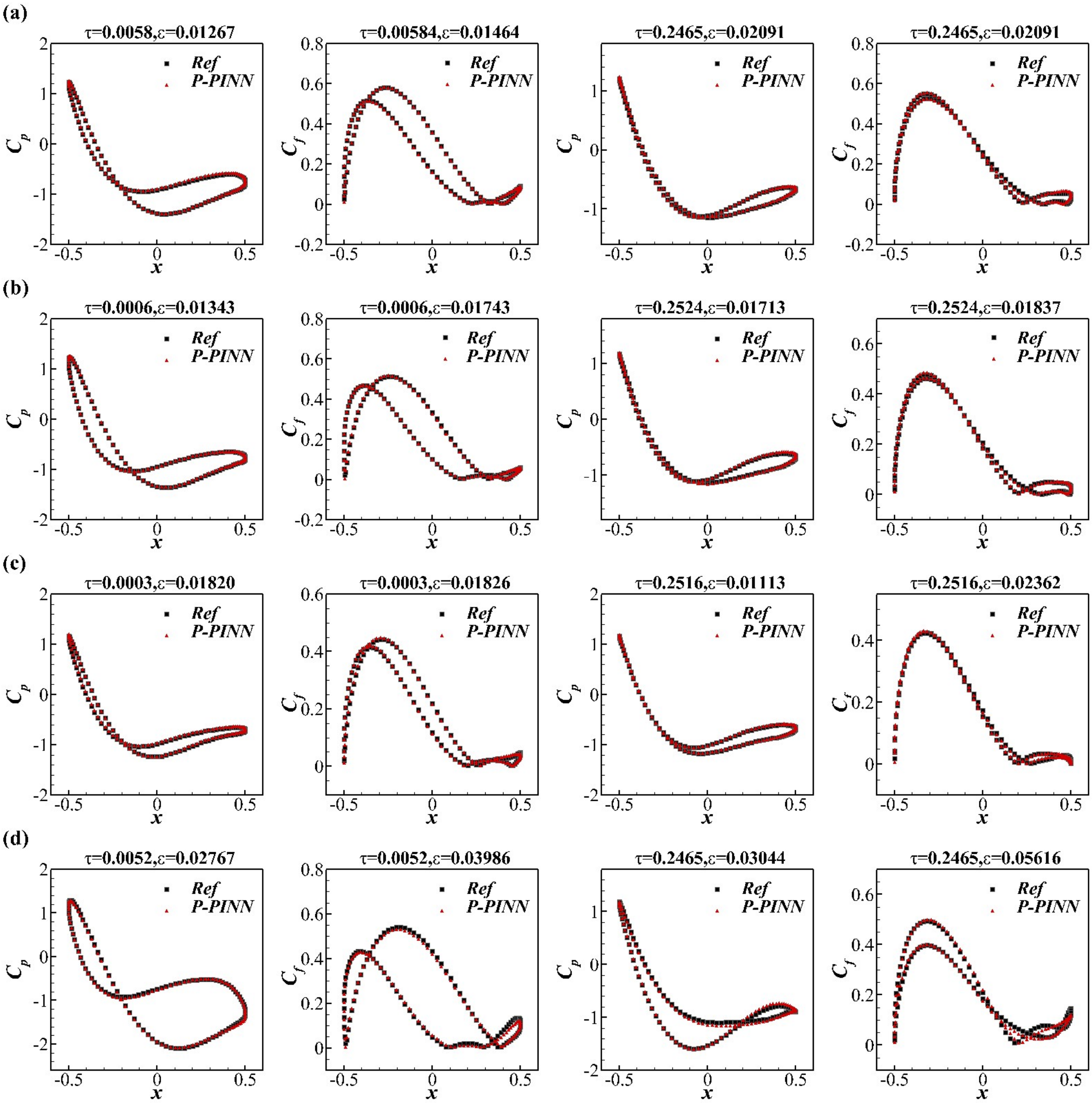

Figure 12. Comparison of $C_p$ and $C_f$ between P-PINN and Fluent for the four validation cases at two representative phases, $\tau \approx 0$ and $\tau \approx 0.25$: (a) Case 1; (b) Case 2; (c) Case 3; (d) Case 4.

To further validate the capability of the model to represent wake flows at supercritical Reynolds numbers, Figure 13 compares the instantaneous vorticity fields of the four validation cases at phase close to $\tau = 0$ with the corresponding Fluent results. Compared with the preceding subcritical cases, the wake vortices at supercritical Reynolds numbers are significantly stronger, and the periodic vortex-shedding structures are more distinct. As shown, P-PINN accurately captures the locations, shapes, and intensities of the near-wall shear layers and downstream wake vortices around the cylinder. The wake structures under different parameter conditions agree well with the Fluent results, and the pointwise errors remain relatively low. These results further indicate that the local deviation in the weak-wake region of subcritical Case 1 does not originate from a general deficiency of P-PINN in representing periodic vortex-shedding structures. For the supercritical cases with more pronounced vortex structures, the parametric P-PINN still maintains high flow-field resolution accuracy.

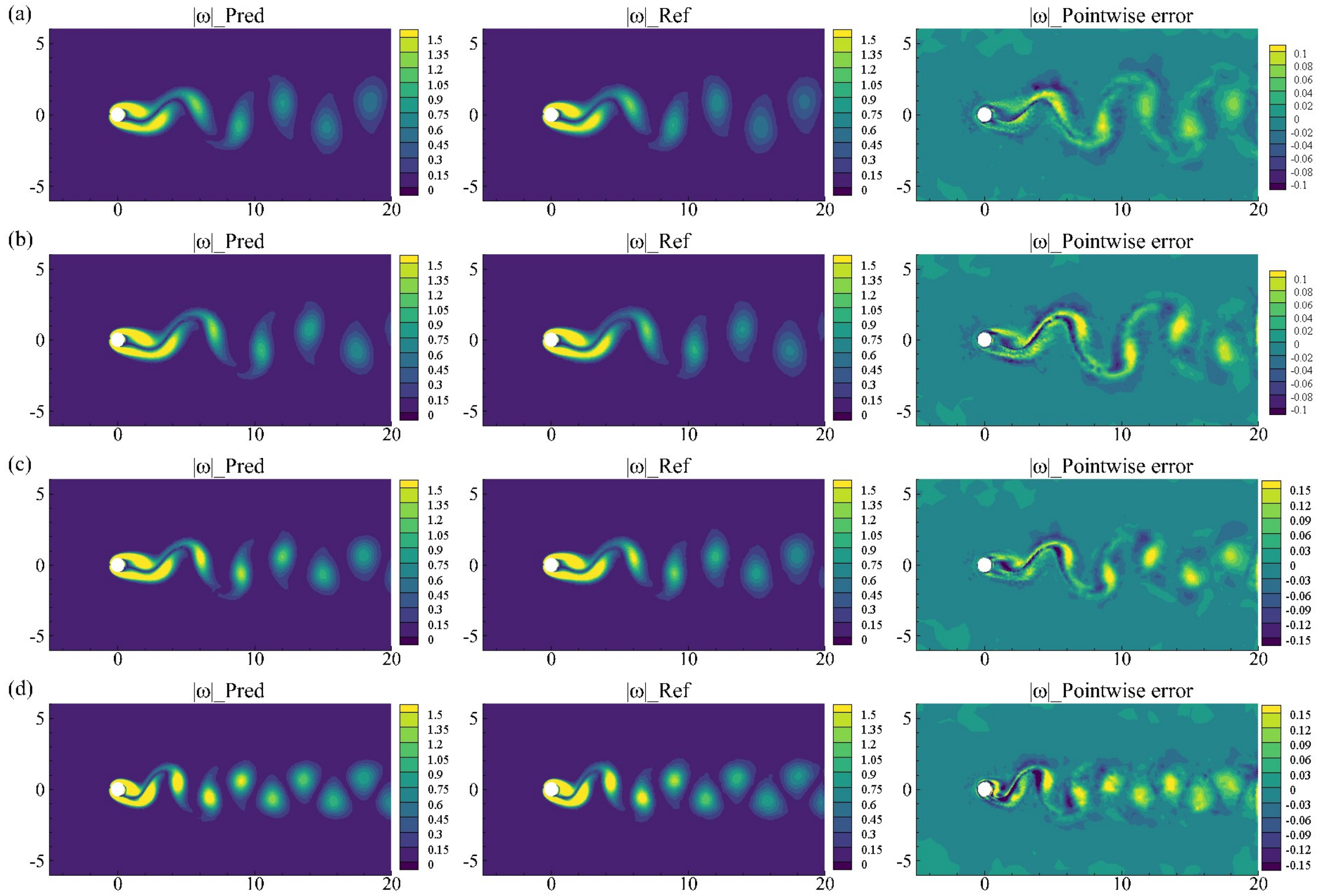

Figure 13. Comparison of the vorticity magnitude $|\omega|$ between P-PINN and Fluent at phases $\tau \approx 0$ for the four validation cases, together with the corresponding pointwise errors: (a) Case 1; (b) Case 2; (c) Case 3; (d) Case 4.

Overall, although the small-amplitude $C_L$ is sensitive to local load errors, the high accuracy of $C_p$ and $C_f$ at different phases, together with the good agreement between the wake flow fields and the Fluent results, demonstrates that the parametric P-PINN can accurately obtain the complete-cycle surface-pressure and skin-friction distributions, drag response, and wake-vortex structures under different supercritical lock-in states. Compared with evaluating solution accuracy solely through integrated aerodynamic forces, these flow-field and surface-load results provide more direct evidence that P-PINN can accurately describe the physical characteristics and flow structures of periodic flows under different parameter conditions. Combined with the preceding subcritical Reynolds-number results, these findings demonstrate that the proposed parametric model maintains good solution accuracy across different Reynolds numbers, motion parameters, and wake characteristics, thereby validating the capability of P-PINN to solve periodic flows parametrically over continuous parameter spaces.

### 3.3 Parametric Oscillating Airfoil

Building on the cylinder cases, we further consider a NACA 0012 airfoil to investigate the capability of P-PINN to solve parametric periodic flows with different geometries and motion types. Unlike the viscous cylinder flows considered above, the flow around the airfoil is described by the incompressible Euler equations, and two representative forced motions, pitching and plunging, are considered. Because viscous shear stress is absent at the wall under inviscid conditions, the parametric solutions are primarily validated in terms of the lift coefficient $C_L$, surface pressure coefficient $C_p$, and velocity components $u$ and $v$. For each motion type, four parameter states are selected and compared with the corresponding Fluent results.

#### 3.3.1 Pitching Motion

We first consider the forced-pitching motion of the NACA 0012 airfoil, with the parameter ranges set to $\alpha \in [-2,5], A \in [1,6], U^* \in [2,7]$. Table 3 lists the specific parameters of the four validation cases and the corresponding RRMSEs of $C_L$ and $C_p$. Figure 14 compares the variations of $C_L$ obtained by P-PINN and Fluent over one motion cycle for the four validation cases. Because the pitching motion is affected by the angle of attack, the surface-pressure distribution no longer exhibits simple half-cycle symmetry. Therefore, four representative phases close to $\tau = 0$, 0.25, 0.5, and 0.75 are selected in Figure 15 to compare the $C_p$ distributions over the airfoil surface.

Table 3. Parameter settings of the validation cases and corresponding RRMSE of $C_L$ and $C_p$

| | $\alpha$ | $A$ | $U^*$ | $\varepsilon_C_L$ | $\varepsilon_C_p$ |
|---|---|---|---|---|---|
| **Case 1** | 0 | 5 | 5 | 0.0455 | 0.0669 |
| **Case 2** | -1 | 2 | 6 | 0.0486 | 0.0685 |
| **Case 3** | 2 | 4 | 3 | 0.0508 | 0.0568 |
| **Case 4** | 4 | 3 | 4 | 0.0291 | 0.0480 |

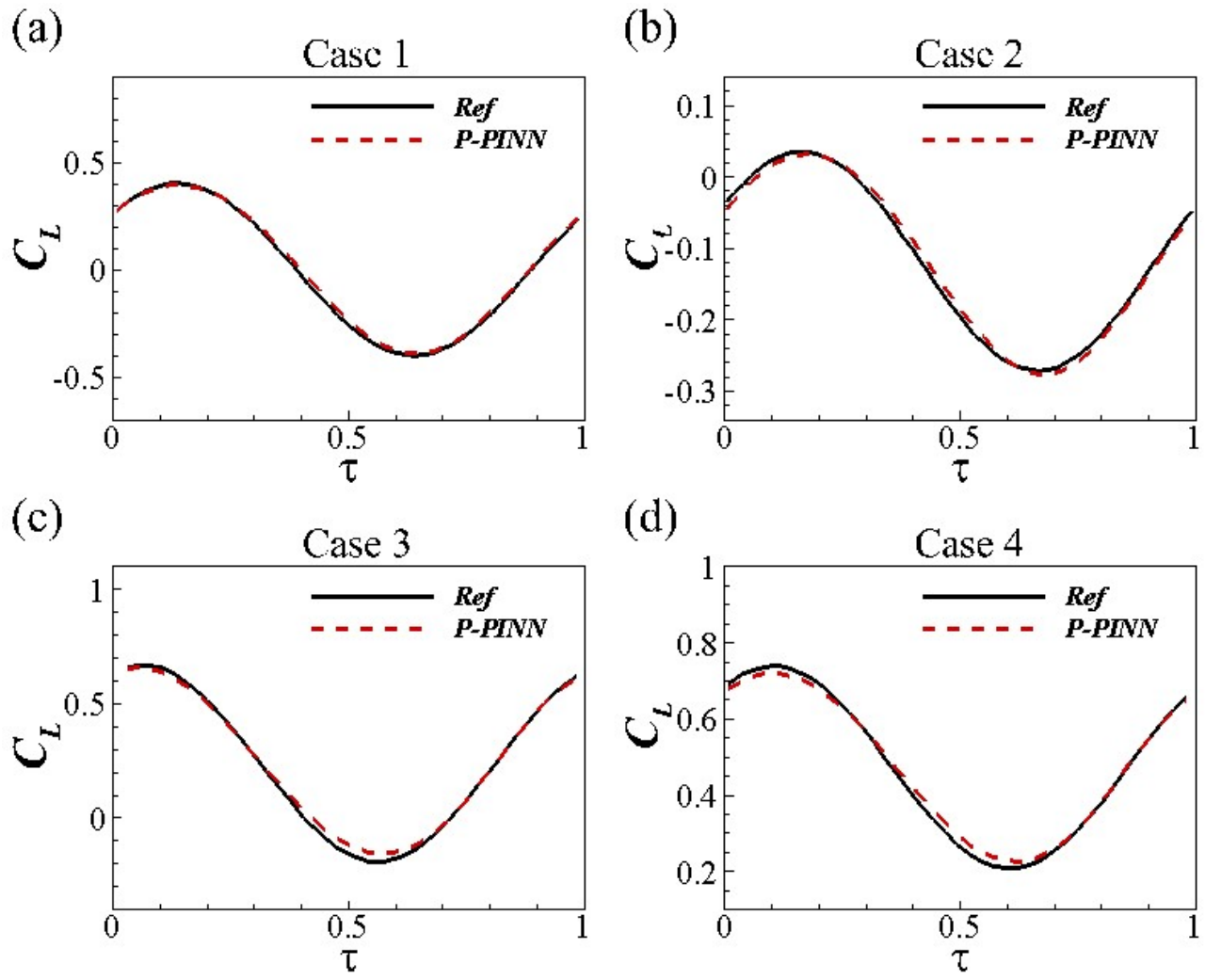

Figure 14. Comparison of the $C_L$ between P-PINN and Fluent for the four validation cases: (a) Case 1; (b) Case 2;

(c) Case 3; (d) Case 4.

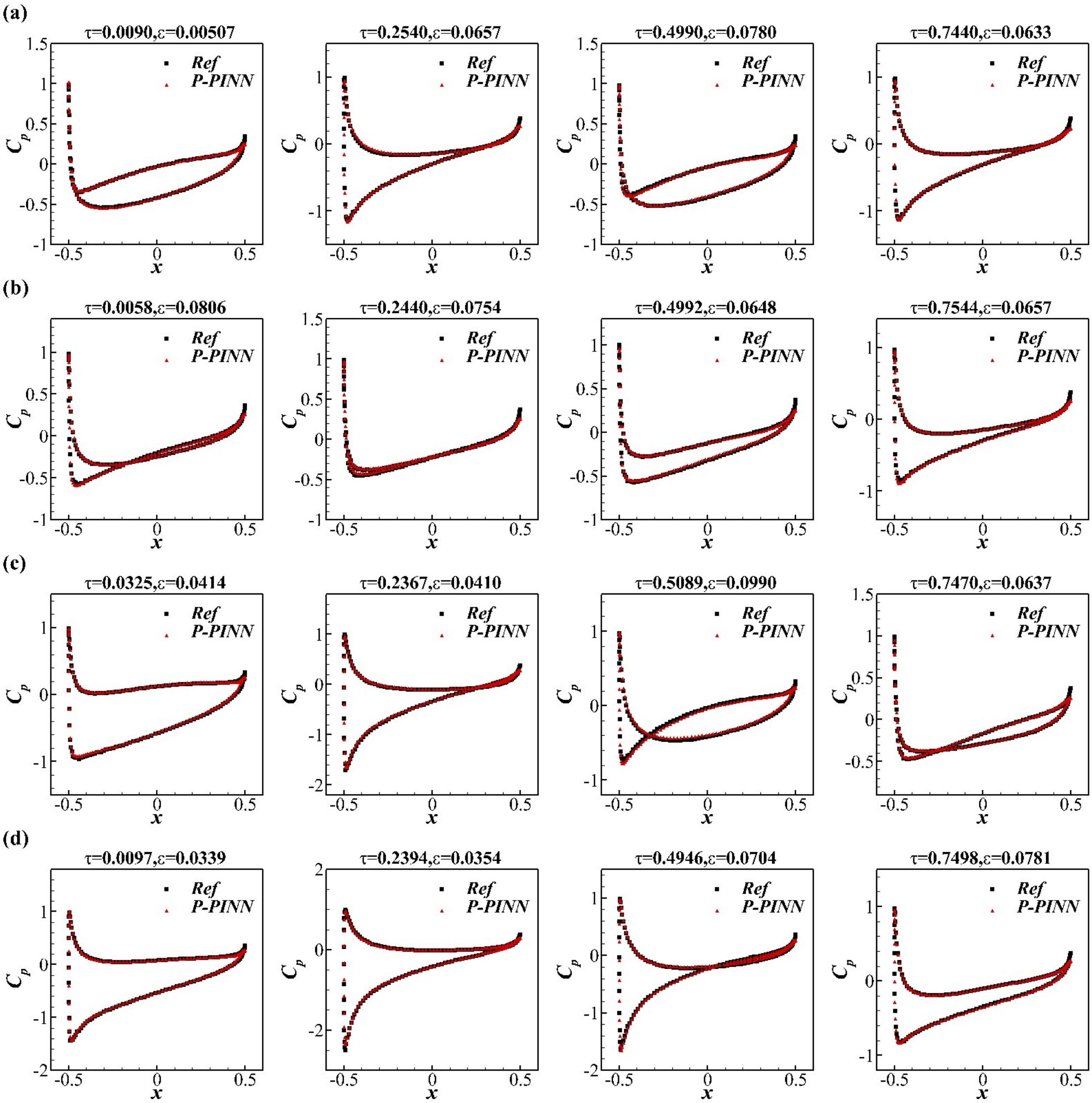

Figure 15. Comparison of $C_p$ between P-PINN and Fluent for the four validation cases at four representative phases, $\tau \approx 0, 0.25, 0.5$ and $0.75$ : (a) Case 1; (b) Case 2; (c) Case 3; (d) Case 4.

The results show that P-PINN accurately captures the periodic lift variations induced by pitching under different freestream conditions and motion parameters. The predicted $C_L$ agrees well with the Fluent results in terms of amplitude, phase, and overall variation trend. Meanwhile, the $C_p$ distributions over the entire airfoil surface at all representative phases are accurately reproduced, including the large pressure gradients near the leading edge and the pressure distributions on the upper and lower surfaces. The error results in Table 3 further demonstrate that the parametric P-PINN maintains high accuracy in predicting both aerodynamic forces and surface pressure under different parameter conditions.

Furthermore, Figure 16 compares the velocity components $u$ and $v$ obtained by P-PINN and Fluent for the four validation cases at phase close to $\tau = 0$. Because the wake-vortex structures are relatively weak in inviscid flow, the instantaneous flow field is primarily evaluated through the velocity distributions. As shown, P-PINN accurately captures the local velocity variations around the airfoil, and the velocity fields under different parameter conditions agree well with the Fluent results, further validating its capability to solve periodic pitching flows parametrically.

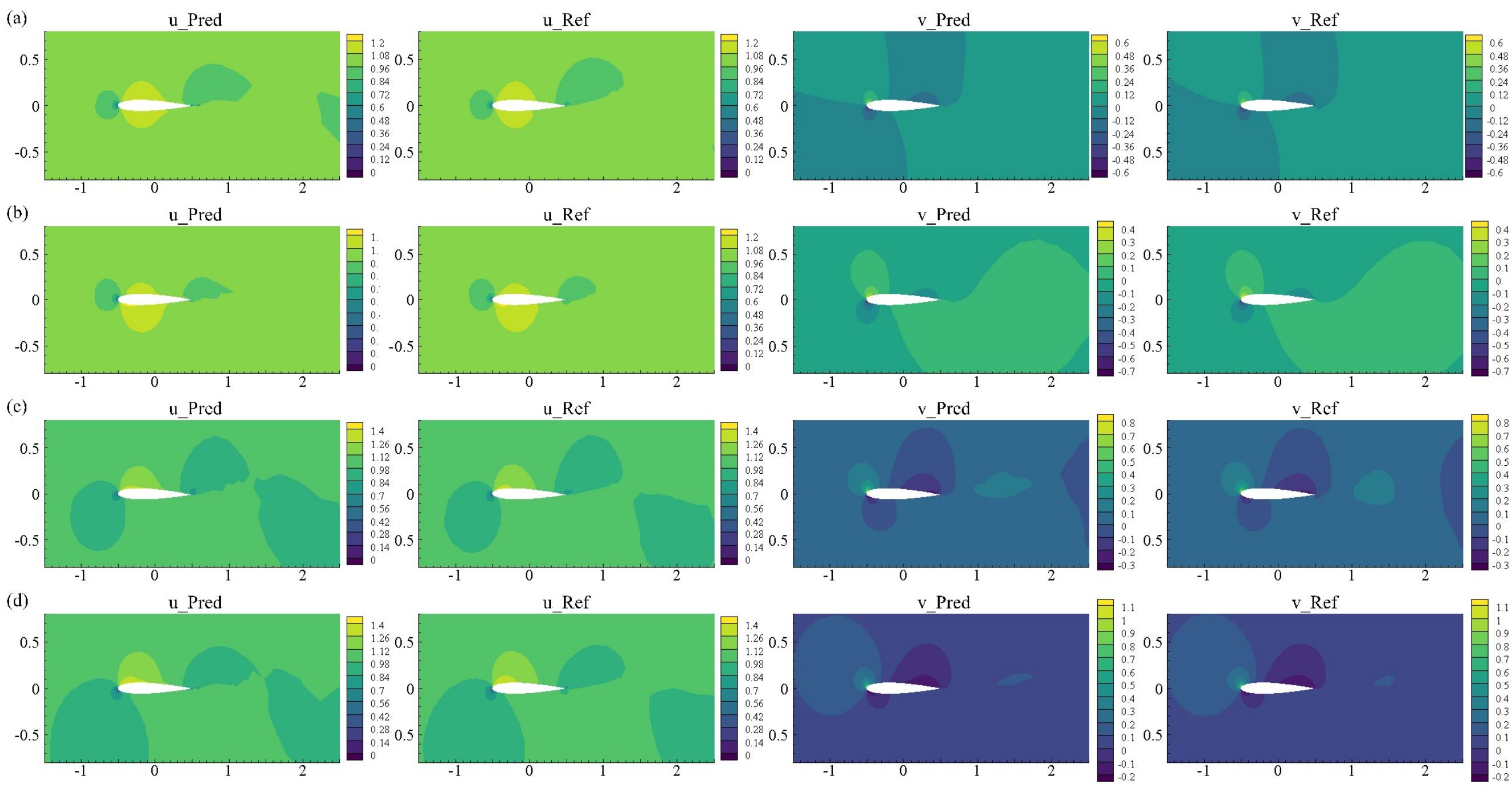

Figure 16. Comparison of the instantaneous velocity components $u$ and $v$ between P-PINN and Fluent at phases $\tau \approx 0$ for the four validation cases: (a) Case 1; (b) Case 2; (c) Case 3; (d) Case 4.

### 3.3.2 Plunging Motion

We then consider the forced-plunging motion of the airfoil, with the parameter ranges set to $\alpha \in [-2,5], A \in [0.02, 0.12], U^* \in [2,7]$. Table 4 summarizes the parameter combinations of the four validation cases and the corresponding RRMSEs of $C_L$ and $C_p$. Figure 17 shows the periodic variations of $C_L$ for the four cases, while Figure 18 compares the $C_p$ distributions over the airfoil surface at four representative phases close to $\tau = 0$, 0.25, 0.5, and 0.75.

Table 4. Parameter settings of the validation cases and corresponding RRMSE of $C_L$ and $C_p$

| | $\alpha$ | ***A*** | ***U**** | $\varepsilon_C_L$ | $\varepsilon_C_p$ |
|---|---|---|---|---|---|
| **Case 1** | 0 | 0.10 | 5 | 0.0667 | 0.0951 |
| **Case 2** | -1 | 0.05 | 6 | 0.0539 | 0.0970 |
| **Case 3** | 2 | 0.06 | 3 | 0.0543 | 0.0805 |
| **Case 4** | 4 | 0.08 | 4 | 0.0321 | 0.0692 |

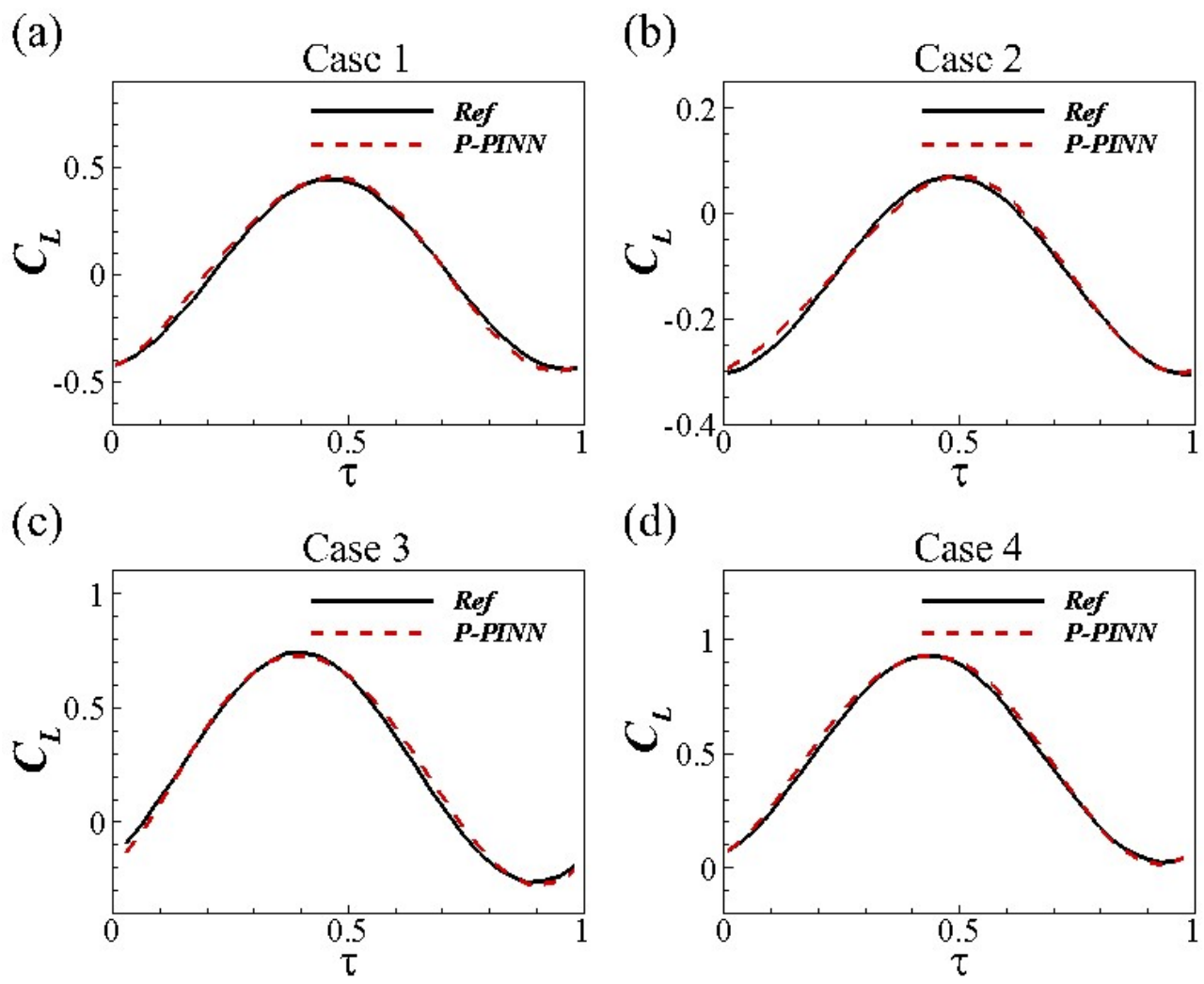

Figure 17. Comparison of the $C_L$ between P-PINN and Fluent for the four validation cases: (a) Case 1; (b) Case 2; (c) Case 3; (d) Case 4.

The results show that the periodic lift responses vary significantly across different motion types, flow conditions and motion parameters, while P-PINN accurately captures their amplitude and phase variations. Meanwhile, the $C_p$ distributions at the representative phases agree well with the Fluent results, accurately reproducing the pressure variations near the leading edge, the pressure difference between the upper and lower surfaces, and the overall chordwise pressure distribution. Together with the RRMSEs in Table 4, these results demonstrate that the parametric P-PINN maintains high accuracy in predicting both the overall aerodynamic forces and local surface loads under different plunging conditions.

Figure 19 further compares the distributions of $u$ and $v$ obtained by P-PINN and Fluent for the four validation cases at phase close to $\tau = 0$. The instantaneous velocity fields predicted by P-PINN agree well with the Fluent results and accurately reflect the local flow variations induced by the plunging motion. The comparisons of $C_L$, $C_p$, and the velocity fields demonstrate that P-PINN also achieves high parametric solution accuracy for forced-plunging airfoil flows.

Overall, for both pitching and plunging motions, P-PINN accurately predicts the periodic lift, surface pressure distribution, and instantaneous velocity field under different flow and motion parameters, demonstrating its capability to solve parametric periodic flows with different motion types.

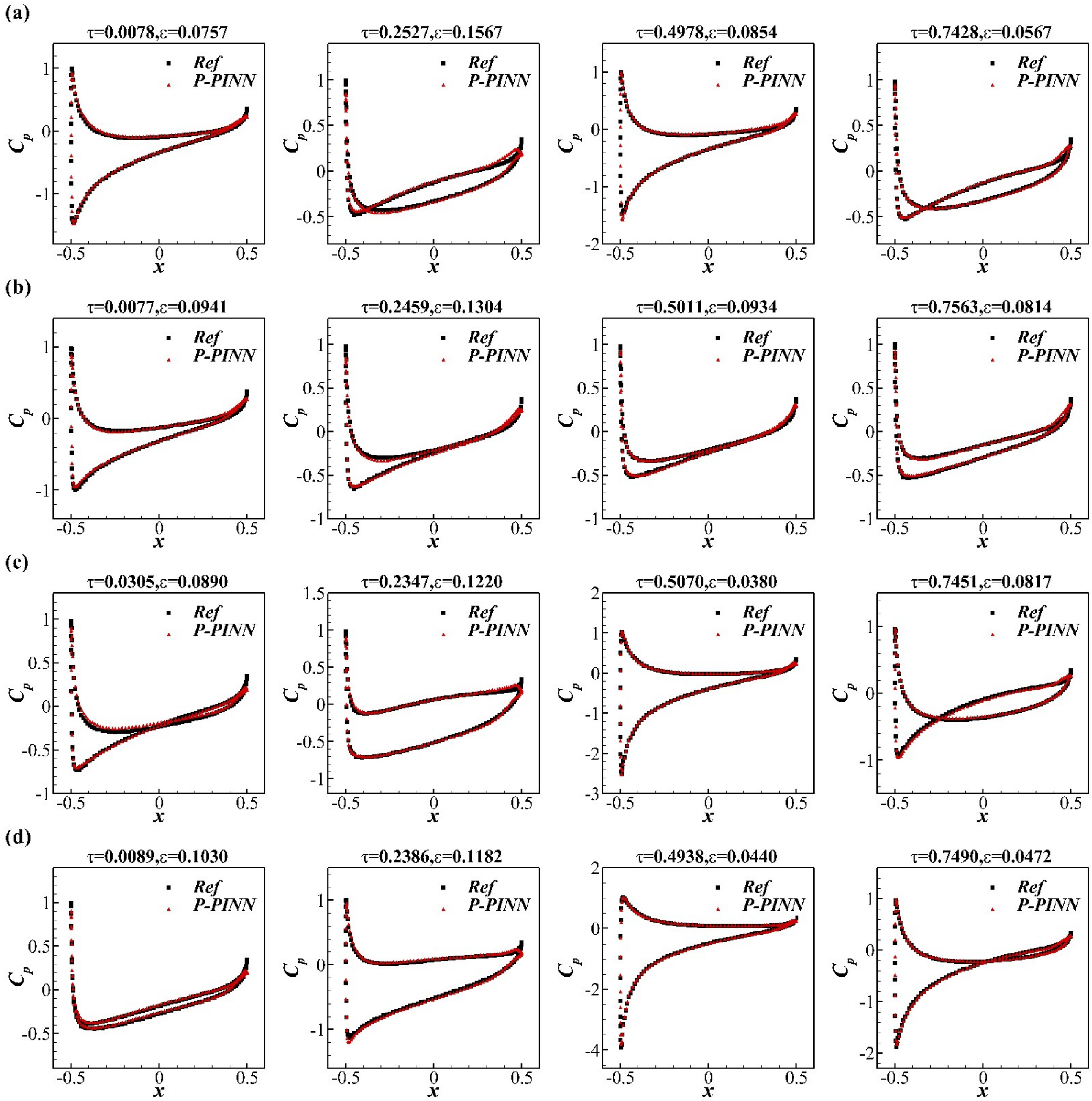

Figure 18. Comparison of $C_p$ between P-PINN and Fluent for the four validation cases at four representative phases, $\tau \approx 0, 0.25, 0.5$ and $0.75$: (a) Case 1; (b) Case 2; (c) Case 3; (d) Case 4.

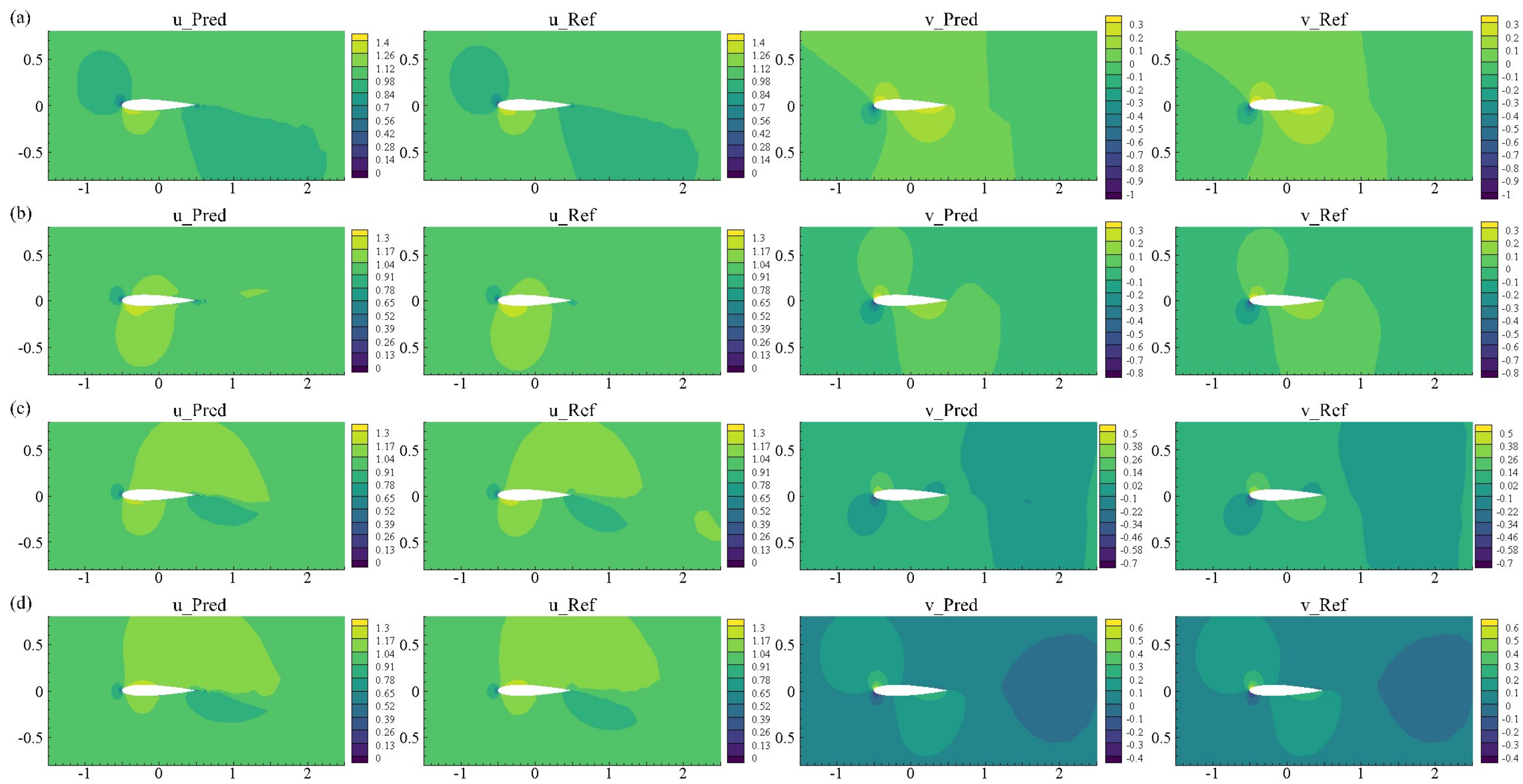

Figure 19. Comparison of the instantaneous velocity components $u$ and $v$ between P-PINN and Fluent at phases $\tau \approx 0$ for the four validation cases: (a) Case 1; (b) Case 2; (c) Case 3; (d) Case 4.

# 4 Nonlinear Aeroelastic Response

Building on the preceding parametric solution of periodic flows, this section further combines P-PINN with the structural dynamics equation to solve finite-amplitude aeroelastic responses with a single degree of freedom. The parametric P-PINN establishes a mapping between the motion amplitude, motion frequency, and aerodynamic forces over a complete period. Therefore, for given structural parameters, the original fluid–structure interaction problem can be transformed into a nonlinear system of equations for the response amplitude and response frequency using first-order harmonic balance. The corresponding aeroelastic response can thus be obtained directly without performing long-time, fully coupled CFD/CSD time marching simulations.

## 4.1 Single-Degree-of-Freedom Aeroelastic Formulation

Consider an elastically mounted circular cylinder in a uniform freestream, with only the transverse plunging degree of freedom retained. Using the nondimensional convective time $t^* = U_\infty t / D$, the structural equation of motion is written as

$$\frac{d^2Y}{dt^{*2}} + 4\pi\zeta F_n \frac{dY}{dt^*} + (2\pi F_n)^2 Y - \frac{2}{\pi M} C_L = 0, \tag{4.1}$$

where, $Y = y / D$ denotes the nondimensional transverse displacement, $M$ and $\zeta$ are the mass ratio and structural damping ratio, respectively, and $F_n$ is the nondimensional structural natural frequency. The corresponding structural reduced velocity is defined as:

$$U_n^* = \frac{U_\infty}{f_n D} = \frac{1}{F_n}. \tag{4.2}$$

This study considers single-degree-of-freedom finite-amplitude periodic states dominated by the fundamental harmonic and adopts the following first-order harmonic approximation:

$$Y \approx A \sin\theta, \qquad \theta = 2\pi F_r t^*, \tag{4.3}$$

where $A$ is the response amplitude and $F_r$ is the nondimensional frequency of the actual response. The corresponding response reduced velocity is defined as:

$$U_r^* = \frac{U_\infty}{f_r D} = \frac{1}{F_r}. \tag{4.4}$$

The lift coefficient predicted by P-PINN over one complete period is expanded as:

$$C_L = \overline{C}_L + C_{L,s} \sin(2\pi F_r t^*) + C_{L,c} \cos(2\pi F_r t^*) + \cdots, \tag{4.5}$$

where:

$$C_{L,s} = \frac{1}{\pi} \int_0^{2\pi} C_L(\theta) \sin\theta d\theta, \qquad C_{L,c} = \frac{1}{\pi} \int_0^{2\pi} C_L(\theta) \cos\theta d\theta. \tag{4.6}$$

Substituting the structural displacement (4.3) and the expansions of periodic lift (4.5) into Eq. (4.1), and balancing the sine and cosine terms separately, gives the first-order harmonic-balance equations:

$$C_{L,s} = 2\pi^3 MA(F_n^2 - F_r^2) = 2\pi^3 MA\left[\left(\frac{1}{U_n^*}\right)^2 - \left(\frac{1}{U_r^*}\right)^2\right], \tag{4.7}$$

$$C_{L,c} = 4\pi^3 M\zeta F_n F_r A = 4\pi^3 M\zeta A\left(\frac{1}{U_n^*}\right)\left(\frac{1}{U_r^*}\right). \tag{4.8}$$

Equation (4.7) describes the effect of the lift component in phase with the displacement on the equivalent stiffness and response frequency of the system, whereas Eq. (4.8) describes the balance between the aerodynamic energy input associated with the velocity-in-phase component and the energy dissipated by structural damping.

The parametric P-PINN established above provides a mapping between the motion parameters and the aerodynamic forces over a complete period. Therefore, for a given $(A, U_r^*)$, $C_L = C_L(\theta; A, U_r^*)$ can be obtained directly. The corresponding $C_{L,s}$ and $C_{L,c}$ can then be calculated from the periodic lift predicted by P-PINN, and two harmonic-balance residuals are defined as:

$$R_s(A,U_r^*) = C_{L,s}(A,U_r^*) - 2\pi^3 MA\left[\left(\frac{1}{U_n^*}\right)^2 - \left(\frac{1}{U_r^*}\right)^2\right],$$
$$R_c(A,U_r^*) = C_{L,c}(A,U_r^*) - 4\pi^3 M\zeta A\left(\frac{1}{U_n^*}\right)\left(\frac{1}{U_r^*}\right). \tag{4.9}$$

Thus, for a given structural reduced velocity $U_n^*$, the single-degree-of-freedom aeroelastic response can be transformed into the following two-dimensional nonlinear system:

$$R_s(A,U_r^*) = 0, \qquad R_c(A,U_r^*) = 0. \tag{4.10}$$

It should be noted that, for the single-degree-of-freedom aeroelastic problem considered here, the zero-amplitude state is a trivial solution of the nonlinear system. To search for the finite-amplitude response branch, the lower bound of the amplitude parameter in the parametric model is used as the minimum allowable response amplitude in the actual calculations. In addition, the converged solution at the preceding structural reduced velocity is used as the initial value for the solution at the next structural reduced velocity. This procedure prevents the solution from converging to the zero-amplitude state.

By simultaneously solving the above bounded nonlinear optimization problems, the corresponding finite-amplitude response $A$ and actual response frequency $F_r = 1/U_r^*$ can be obtained. Repeating the solution procedure for different $U_n^*$ values further obtains the nonlinear aeroelastic response curves of the response amplitude and response frequency as functions of the structural reduced velocity.

**4.2 Aeroelastic Response Results**

To provide an initial validation of the proposed aeroelastic solution method, the parametric P-PINN is retrained for cylinder flow at a subcritical Reynolds number. The Reynolds number is fixed at $Re = 33$, and the parameter ranges of the motion amplitude and response reduced velocity are set to $A \in [0.05, 0.5], U_r^* \in [4,11]$, respectively. Four parameter states are selected within this range to validate the accuracy of the periodic aerodynamic-force predictions. The specific parameters and the RRMSEs of the force coefficients are listed in Table 5. Figure 20Figure 21present comparisons of the periodic lift coefficients and vorticity fields at phase close to $\tau = 0$, respectively. The results show that P-PINN accurately captures the amplitude and phase variations of $C_L$ under different motion amplitudes and response frequencies. It also accurately reproduces the locations, shapes, and intensities of the near-wall shear layers and wake vortices. These results demonstrate that the parametric model provides high accuracy in predicting both periodic aerodynamic forces and instantaneous flow fields, thereby providing a reliable basis for the subsequent solution of the aeroelastic response.

Table 5. Parameter settings of the validation cases and corresponding RRMSE of $C_L$, $C_D$, $C_p$ and $C_f$

| | ***Re*** | ***A*** | ***U**** | $\varepsilon_C_L$ | $\varepsilon_C_D$ | $\varepsilon_C_p$ | $\varepsilon_C_f$ |
|---|---|---|---|---|---|---|---|
| **Case 1** | 33 | 0.05 | 5 | 0.1063 | 0.0019 | 0.0200 | 0.0092 |
| **Case 2** | 33 | 0.25 | 6 | 0.0331 | 0.0060 | 0.0413 | 0.0137 |
| **Case 3** | 33 | 0.45 | 7 | 0.0621 | 0.0108 | 0.0446 | 0.0146 |
| **Case 4** | 33 | 0.10 | 10 | 0.1379 | 0.0025 | 0.0170 | 0.0086 |

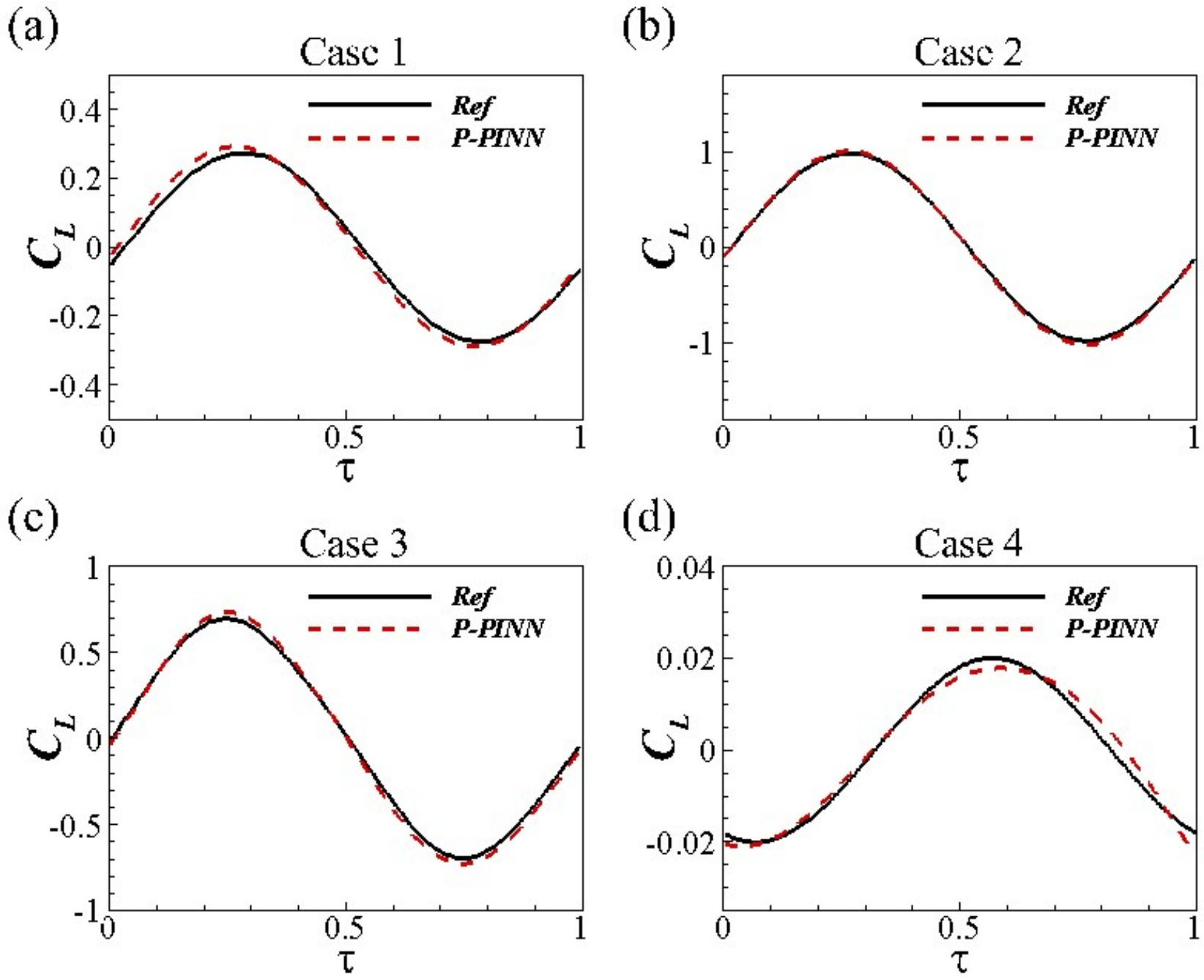

Figure 20. Comparison of the $C_L$ between P-PINN and Fluent for the four validation cases: (a) Case 1; (b) Case 2; (c) Case 3; (d) Case 4.

Building on this, the previously obtained parametric model is further used to solve the single-degree-of-freedom aeroelastic response of the circular cylinder. The calculation parameters are set to $Re = 33, M = 9.407, \zeta = 0$ . The first-order harmonic-balance equations are solved for different structural reduced velocities $U_n^*$, yielding the corresponding response amplitudes $A$ and response reduced velocities $U_r^*$, or equivalently, the response frequencies $F_r$. To validate the results, the aeroelastic responses obtained using P-PINN and the harmonic-balance method are compared with the fluid–structure interaction results from Fluent. Figure 22 presents the variations of the response amplitude and response frequency with the structural reduced velocity.

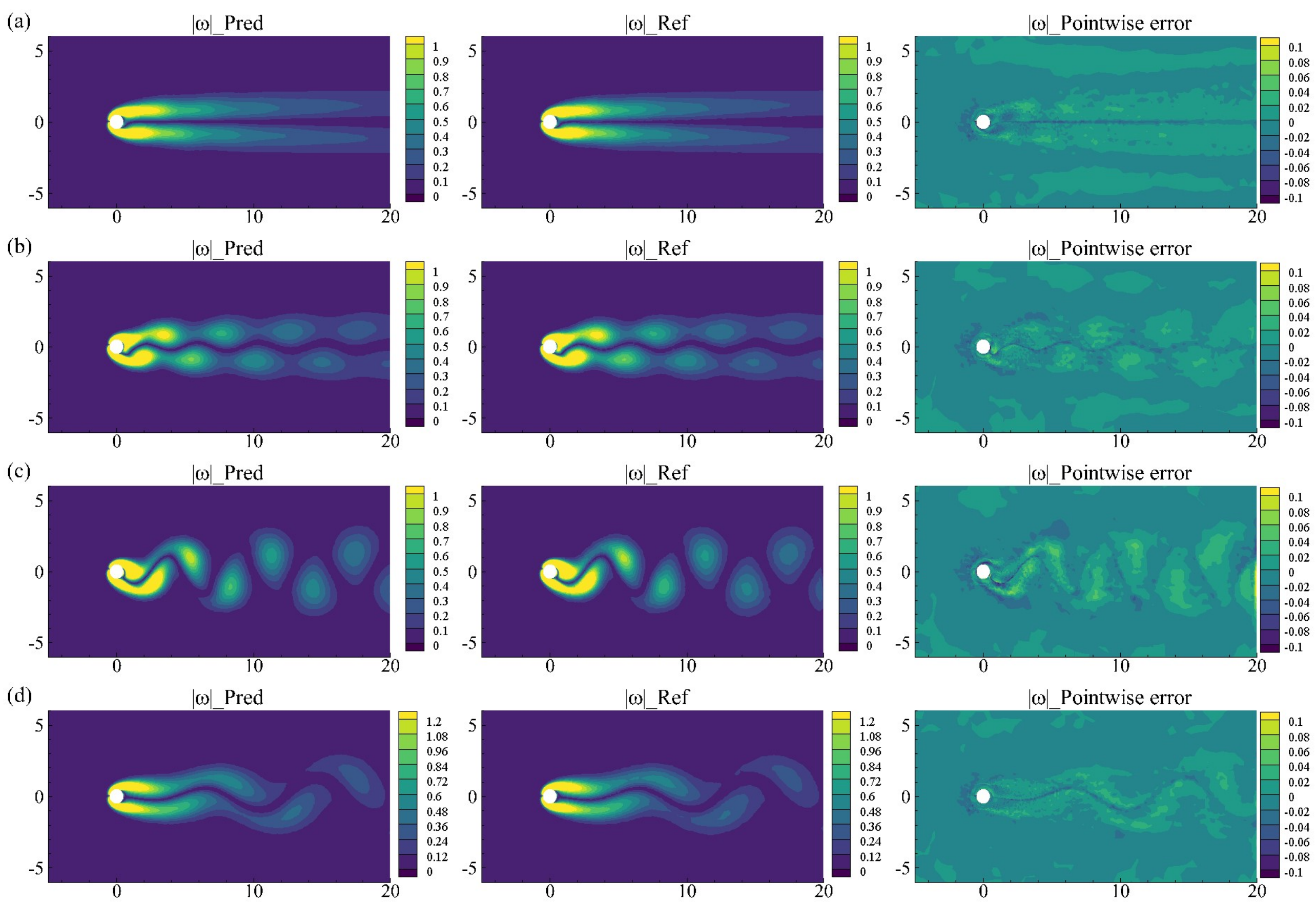

Figure 21. Comparison of the vorticity magnitude $|\omega|$ between P-PINN and Fluent at phases $\tau \approx 0$ for the four validation cases, together with the corresponding pointwise errors: (a) Case 1; (b) Case 2; (c) Case 3; (d) Case 4.

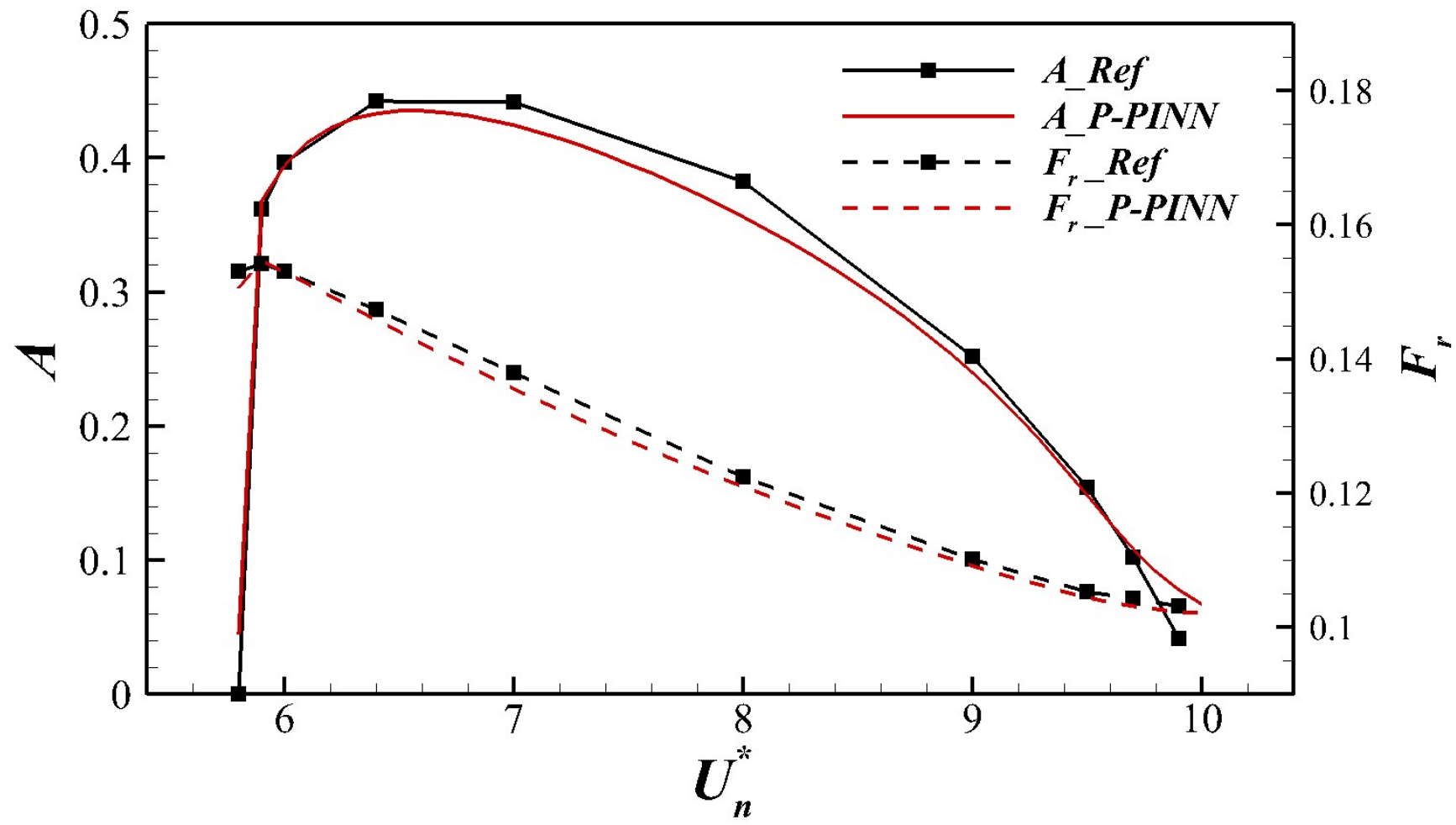

Figure 22. Comparison of the aeroelastic response curves obtained by P-PINN and Fluent as functions of the structural reduced velocity $U_n^*$, including the response amplitude and response frequency.

As shown in Figure 22, the amplitude-response curve obtained by P-PINN accurately reproduces the overall aeroelastic response process, including the onset of oscillation, rapid amplitude growth, attainment of the maximum response, and subsequent gradual decay. The response amplitudes and frequencies in the onset region, near the maximum response, and in the decay region agree well with the Fluent results. In particular, near the onset of oscillation, the response amplitude varies rapidly with $U_n^*$, yet P-PINN still accurately captures the steep variation in the response curve. Meanwhile, $F_r$ exhibits local variations in this region that differ from the overall decreasing trend, and these variations are also accurately reproduced. Overall, the proposed method captures the main features of the response amplitude and response frequency as the structural parameters vary.

Despite the good overall agreement, some amplitude deviation remains in the relatively large-$U_n^*$ region near the termination of oscillation. This deviation is mainly associated with the small response amplitude in this region and the fact that the corresponding states are close to the boundary of the training parameter domain of the parametric P-PINN. Taking validation Case 4 as an example, when $A = 0.1$ and $U_r^* = 10$, the amplitude of $C_L$ is only approximately 0.02. Such a small aerodynamic-force amplitude makes the prediction more sensitive to local errors, which further affects the response amplitude obtained from the harmonic-balance equations. Nevertheless, the proposed method still accurately captures the overall gradual decrease in response amplitude near the termination of oscillation, and the response frequency remains in good agreement with the Fluent results. Overall, the combination of P-PINN and first-order harmonic balance accurately obtains the finite-amplitude aeroelastic response of the single-degree-of-freedom cylinder and maintains high solution accuracy over most of the response range.

In addition, because the periodic aerodynamic forces required during the aeroelastic solution are directly provided by the trained parametric P-PINN, the complete aeroelastic response curve can be obtained within only a few seconds once the structural parameters are specified. When the structural mass, damping, or other parameters are further varied, the corresponding response amplitudes and frequencies can be obtained rapidly without resolving the flow field or performing long-time coupled CFD/CSD time marching. At present, the main computational cost of the method remains concentrated in the offline training of the parametric P-PINN, with the training of a single model typically requiring several days. Therefore, its main advantage lies in enabling rapid online aeroelastic analysis for a large number of structural parameter states after a single offline training process.

Overall, by using the complete-cycle aerodynamic forces provided by the parametric P-PINN and combining them with the first-order harmonic-balance method, this study achieves the direct solution of finite-amplitude periodic aeroelastic responses of a single-degree-of-freedom cylinder. The method simultaneously obtains the response amplitude and actual response frequency at different structural reduced velocities, and the results agree well with the Fluent fluid–structure interaction simulations in the onset, maximum-amplitude, and decay regions. These results validate its ability to accurately represent the variation of the nonlinear aeroelastic response. Although the offline training cost of the current parametric model remains relatively high, once trained, the model enables rapid evaluation of aeroelastic responses over a wide range of structural parameters without repeated fluid–structure interaction time marching. This capability establishes a new and efficient framework for large-scale parametric analysis of periodic nonlinear aeroelastic responses.

## 5 Conclusions and Outlook

This study proposes a physics-driven framework for solving parametric periodic flows with moving boundaries and finite-amplitude aeroelastic responses exhibiting stable periodic behavior. The Periodic Physics-Informed Neural Network (P-PINN) introduces temporal periodicity over a single motion cycle, transforming the conventional long-time initial-value problem into a temporal boundary-value problem over one cycle. This enables direct solution of periodic unsteady flows without resolving the preceding transient evolution toward the periodic state. By further introducing flow and motion parameters as network inputs, the P-PINN establishes a continuous mapping from the parameter space to the periodic flow field within a single model. The parametric aerodynamic model is then coupled with the structural dynamic equation through first-order harmonic balance, transforming the single-degree-of-freedom fluid–structure interaction problem into a low-dimensional nonlinear system with the response amplitude and response frequency as the unknowns, thereby enabling direct solution of finite-amplitude periodic aeroelastic responses.

The proposed method was validated through forced plunging motion of a circular cylinder and forced plunging and pitching motions of an NACA 0012 airfoil, covering the incompressible Navier–Stokes and Euler equations under different flow conditions, motion amplitudes, and reduced velocities. The results show that P-PINN can accurately reproduce the periodic aerodynamic forces, surface loads, and instantaneous flow fields over the investigated parameter ranges, while establishing a continuously callable parametric model of periodic flow. Furthermore, for an elastically supported circular cylinder at subcritical Reynolds numbers, the parametric periodic aerodynamic model was coupled with the structural dynamic equation. The resulting response amplitude and response frequency as functions of the structural reduced velocity show overall good agreement with the fully coupled Fluent fluid–structure interaction results, demonstrating that the proposed framework can further obtain free aeroelastic periodic responses from periodic aerodynamic information established over the prescribed-motion parameter space.

An important feature of the proposed framework is that the parametric aerodynamic model requires only a single offline training, after which the trained model can be repeatedly evaluated for different flow and motion parameter states. On this basis, the parametric aerodynamic model can be further coupled with the structural dynamic equation to obtain aeroelastic response amplitudes and frequencies under different structural parameters without repeatedly performing long-time fluid–structure interaction simulations. For the cases considered in this study, the complete aeroelastic response curve can be obtained online within several seconds. Therefore, the proposed P-PINN framework enables not only direct solution of parametric periodic flows, but also provides a common aerodynamic basis for subsequent parametric aeroelastic analysis. The current method is primarily intended for unsteady flows with stable periodic behavior, while the aeroelastic analysis further assumes a single-degree-of-freedom response dominated by the fundamental harmonic. Its applicability therefore depends on the periodic-response assumption and the coverage of the relevant flow and motion states by the parametric flow model.

Future work will extend the covered flow, geometric, and motion parameter spaces and apply the framework to more complex aeroelastic systems. For multi-degree-of-freedom problems, motion amplitudes of different degrees of freedom, the common response frequency, and their relative phase can be incorporated into the parametric periodic flow model and coupled with the corresponding multi-degree-of-freedom structural dynamics. For periodic responses with significant higher harmonics, multi-harmonic representations and the corresponding structural balance formulations can be further developed. In addition, maintaining the accuracy and training efficiency of the parametric flow model in higher-dimensional parameter spaces, as well as handling more complex periodic branches and response stability problem, remain important directions for future research.

## Conflict of Interest Statement

The authors have no conflicts to disclose.

## Acknowledgments

We would like to acknowledge the support of the National Natural Science Foundation of China (Grant No.92152301) and the Science and Technology Plan Project of Shaanxi Province (Grant No.2024JC-YBQN-0010).